\documentclass[prb,twocolumn,showpacs]{revtex4-1}
\usepackage{amsmath,amssymb,mathrsfs,bbold}
\usepackage{graphicx,color}
\usepackage{bbm}
\usepackage{epstopdf}

\newcommand{\ket}[1]{\left|#1\right\rangle}
\newcommand{\bra}[1]{\left\langle#1\right|}

\newcommand{\dunderline}[1]{\underline{\underline{#1}}}

\allowdisplaybreaks[4]

\begin{document}
\title{Functional renormalization group in Floquet space}
\author{Anna Katharina Eissing}
\affiliation{Institut f\"ur Theorie der Statistischen Physik, RWTH Aachen University,
52056 Aachen, Germany}
\affiliation{JARA-Fundamentals of Future Information Technology, 52056 Aachen, Germany}
\author{Volker Meden}
\affiliation{Institut f\"ur Theorie der Statistischen Physik, RWTH Aachen University,
52056 Aachen, Germany}
\affiliation{JARA-Fundamentals of Future Information Technology, 52056 Aachen, Germany}
\author{Dante Marvin Kennes}
\affiliation{Institut f\"ur Theorie der Statistischen Physik, RWTH Aachen University,
52056 Aachen, Germany}
\affiliation{JARA-Fundamentals of Future Information Technology, 52056 Aachen, Germany}
\affiliation{Department of Physics, Columbia University, New York, NY, 10027, USA}

\date{\today}
\pagestyle{plain}

\begin{abstract}
We present an extension of the functional renormalization group to Floquet space, which enables us to treat the long time behavior of interacting time periodically driven quantum dots. It is one of its strength that the method is neither bound to small driving amplitudes nor to small driving frequencies, i.e.\,very general time periodic signals can be considered. It is applied to the interacting resonant level model, a prototype model of a spinless, fermionic quantum dot. The renormalization in several setups with different combinations of time periodic parameters is studied, where the numerical results are complemented by analytic expressions for the renormalization in the limit of small driving amplitude. We show how the driving frequency acts as an infrared cutoff of the underlying renormalization group flow which manifests in novel power laws.
We utilize the tunability of the effective reservoir distribution function in a periodically driven onsite energy setup to show how its shape is directly reflected in the renormalization group flow. This allows to flexibly tune the power-law renormalization generically encountered in quantum dot structures. Finally, an in-phase quantum pump as well as a single parameter pump are investigated in the whole regime of driving frequency, demonstrating that the new power law in the driving frequency is reflected in the mean current of the latter.
\end{abstract}

\maketitle

\section{Introduction}

Lately low dimensional systems have been the subject of extensive research both theoretically as well as experimentally. In particular, quantum dots play a central role as they are considered as elementary building blocks for possible quantum information processing devices.\cite{Hansonreview07} 

One active field of research concentrates on setups with time periodically varying external fields with a driving frequency $\Omega$.\cite{Platero2004, Roche13}
A possible application is the quantum pump, where charge is transported by periodic oscillations of the confining potentials without any applied bias voltage. Thouless suggested an analogon of the classical peristaltic pump such that by an adiabatic variation of the external fields, where the driving frequency is small compared to all other energy scales in the system, quantized particle transport can be observed.\cite{Thouless1983}
This has stimulated further theoretical\cite{Brouwer98} and experimental\cite{Switkes99} research of pumping setups in this limit as well as beyond it. \cite{Kaestner2015, Roche13}
The quantization of the transported charge $n e$ has then led to the idea of a single electron pump \cite{Pothier1992, Kaestner2015} as well as single electron currents which could be used as a new standard of the current.\cite{Pekolareview13}

Various methods have been deviced to describe time periodic quantum dot systems, which are often bound to a certain regime of driving frequency. There is a broad range of methods avalaible in the adiabatic limit in which $\Omega$ can be taken as a small parameter \cite{Aleiner98, Splettstoesser2006a, Moskalets2004b, Riwar2013} as well as in the anti-adiabatic limit of $\Omega$ large compared to all other energies.\cite{Braun2008} Methods with which it is possible to access the whole range of driving frequency are often limited to the non-interacting regime. \cite{Moskalets2002}
The regime of driving frequencies of the same order as the dot coupling $\Gamma$ shows interesting effects such as single parameter pumping and reversion of the pump direction.\cite{Citro2003, Cavaliere2009, Braun2008, Croy2012b,  Kashuba2012, Kaestner2008}

One of the main challenges, when describing these systems theoretically is to treat the local on-dot Coulomb interaction of the many-particle systems.\cite{Citro2003, Splettstoesser2006a, Winkler2013, Hiltscher2010, Hiltscher2011, Hernandez2009, Suzuki2015}

We propose an approach to treat interactions in time periodically quantum dot systems without any restrictions concerning driving frequency or amplitude. The influence of the time periodically varied external fields is modeled as usually by time dependent parameters $p(t)$ (referred to as `signal' in the following)  in a microscopic model.\cite{Brouwer98, Splettstoesser2005, Splettstoesser2006a, Hiltscher2010, Hernandez2009, Haupt2013, Cavaliere2009, Croy2012a, Kashuba2012, Moskalets2002, Kohler2005} The shape of the signal can be of arbitrary form in our approach.
We use the functional renormalization group (FRG) method to treat interactions, which has proven to be a versatile tool for interacting low-dimensional systems.\cite{Metzner2012}
It has been extended recently to treat quantum dot system with an explicitly time dependent Hamiltonian.\cite{Kennes2012a}
This approach can be used to study the transient behavior of time periodic systems. The long time dynamics, however, is numerically difficult to access and performing the full time evolution explicitly renders analytical insights of the long time limit out of reach.

 We aim at the steady state of the time periodic system which arises when all transient behavior has died out and the system has inherited the given periodicity. To access this long time behavior directly, we sent the initial time $t_0$ at which the dot and leads are coupled to $-\infty$. 
We set up a FRG procedure, which explicitly takes advantage of the time periodicity of the system making use of the Floquet theorem \cite{Floquet1883} and hence employing Floquet Green's functions. \cite{Tsuji2008, Stefanucci2008, Genske2015} Differently than other methods, which explicitly take advantage of the adiabaticity \cite{Aleiner98, Splettstoesser2006a, Moskalets2004b, Riwar2013}, anti-adiabaticity \cite{Braun2008} or smallness of the driving amplitude our method is not bound to any limit in driving frequency or amplitude. 
The unbiased RG procedure in Floquet space as a natural basis of the time periodic setup, can be applied to various quantum dot setups and can reveal physics arising of the interaction induced correlations including power-law behavior with interaction dependent exponents. A new power law in the driving frequency which is reflected in the mean current of a single parameter pump has already been presented in Ref.\,\onlinecite{Eissing16}. Besides the technical details of the utilized method, we here present further applications and explicitly leave the limit of small driving amplitudes.

This paper is organized as follows.   
In Sec.\,\ref{sec:Green_sec} the general setup is introduced and the Green's function in the Keldysh formalism are defined and transformed to Floquet space.  Furthermore, the observables of interest are defined.  Section \ref{sec:fRG} illustrates the general idea of the functional renormalization group and elaborates on the incorporation of the flow parameter as well as the applied truncation of the hierarchy of differential equations. The time dependent flow equations are then transformed to Floquet space.
Section \ref{sec:model} introduces the interacting resonant level model as an established model of a spinless single level quantum dot and the known physics in the time independent model is presented. Our main results are discussed in Sec.\,\ref{sec:analy}, where we consider the renormalization of the parameters of the model for four different protocols.  Both the hopping matrix elements and/or the onsite energy of the dot are varied periodically in the limit of small driving amplitudes. Besides the numerical solution of the full flow equation, analytical expressions for the renormalized parameters are obtained in the limit of small driving amplitude, which allows to gain a deeper understanding of the underlying renormalization physics and reveal a new power law in the driving frequency. 
This is complemented by the results of Sec.\,\ref{sec:Bessel} for a periodically varied onsite energy in the whole range of driving amplitude, which shows interesting tunable renormalization physics.
In the last Sec.\,\ref{sec:pump} we turn to  quantum pumps as one possible application in the field of periodically driven quantum dot systems. Physical consequences of the afore discussed renormalization are examined. 
The appendix presents the details of the analytic calculations.

\section{Model and Keldysh Green's functions in Floquet space}\label{sec:Green_sec}
\subsection{Hamiltonian} We are aiming at the non-equilibrium transport through periodically driven few level quantum dots
with a Hamiltonian of the form
\begin{equation}
H(t) = \,H_{\text{dot}}(t) +\sum_ {\alpha} \left[ H_{\text{coup},\alpha}(t) + H_{\text{lead,}\alpha} \right], 
\end{equation}
with a general dot Hamiltonian, which consists of a single particle term and a two-particle interaction,
\begin{align}
H_{\text{dot,0}}(t) &=\sum_{ij} \,\epsilon_{ij} (t) d_i^{\dagger} d_j, \\
H_{\text{dot,int}}(t) &=\sum_{ijkl} \, \bar{u}_{ijkl} (t)\, d_i^{\dagger}d_j^{\dagger} d_l d_k \text{ .}
\end{align}
Here $d^{(\dagger)},d$ are annihilation (creation) operators on the dot and the $i,j,k,l$ label the levels of the dot.
The leads are modeled as non-interacting and are tunnel-coupled to the dot 
\begin{align}
 H_{\text{coup},\alpha} (t) &= \sum_{i,q_{\alpha}} v_{q_\alpha,i}(t) \,d_i^{\dagger} c_{q_{\alpha}} + \text{H.c.,}\\
 H_{\text{lead},\alpha}  &=  \sum_{q_{\alpha}}(\epsilon_{q_\alpha} - \mu_{\alpha})\,  c^{\dagger}_{q_{\alpha}} c_{q_{\alpha}},
\end{align} 
with annihilation (creation) operators $c_{q_{\alpha}}^{(\dagger)}$ of reservoir electrons. 
Within our approach any of the parameters indicated by the argument $t$ can be time periodic with the same period $T$. These time periodic variations of the parameters can be of general form; examples are sinusoidal signals and rectangular ones.

\subsection{Green's functions in Keldysh formalism} 
We employ the Keldysh formalism to tackle the non-equilibrium situation with an explicitly time dependent Hamiltonian and thus work with Green's functions depending on two times.\cite{Haug2008, Rammer2007} The Green's functions can be employed to compute transport observables, as e.g.\,the current.\cite{Meir92} The non-interacting reservoirs are projected out resulting in reservoir self-energies, which are discussed later on.
To treat on-dot interactions our approach allows to set up and solve flow equations for the dot one-particle irreducible vertex functions such as the self-energy and effective two-particle interaction. The right hand side of the flow equations depends on the dot Green's function and the vertex functions. The latter are in turn fed back to the Green's function via the Dyson equation, resulting in differential equations to be solved.
We can thus focus on the dot Green's functions with the retarded and Keldysh components defined as
\begin{align}
G^{\text{ret}}_{ij}(t,t') &= -i\, \Theta (t-t') \text{Tr} \rho_0 \{ d_i (t), d_{j}^{\dagger}(t') \}, \\
G^{\rm K}_{ij}(t,t') &= -i\, \text{Tr} \rho_0 [d_i(t), d_j^{\dagger}(t')] ,
\end{align}
with the operators in the Heisenberg picture and $[..,..]$ and $\{..,..\}$ denoting the commutator and the anticommutator, respectively. 
\begin{equation}
\rho_0 = \rho (t=t_0) = \rho^{\rm dot}_0 \otimes \rho^{\rm res}_{\alpha_1,0} \otimes \rho^{\rm res}_{\alpha_2,0} \otimes ... \otimes \rho^{\rm res}_{\alpha_n,0}
\end{equation} 
is the initial density matrix at time $t_0$. The reservoirs are supposed to be in grand canonical equilibrium with temperature $T_\alpha$ and $\mu_\alpha$
\begin{equation}
\rho^{\rm res}_{\alpha,0} = e^{-(H_\alpha- \mu_\alpha N_\alpha)/T_\alpha} / {\rm Tr} \ e^{-(H_\alpha- \mu_\alpha N_\alpha)/T_\alpha},
\end{equation}
with $N_\alpha$ being the particle number operator. We assume that the dot is initially empty and decoupled for $t < t_0$. \\
The advanced Green's function follows as
\begin{equation}
G^{\text{adv}}_{ij} (t,t') = \left[G^{\text{ret}}_{ji} (t',t)\right]^* \text{	.}
\end{equation}

Aiming at time-periodic systems in the long time limit, we will transform the Green's functions to \textit{Floquet space}.\cite{Shirley1965, Wingreen1993, Arrachea2005, Arrachea2006, Wu2008, Tsuji2008, Rentrop2014} This is justified by the \textit{Floquet theorem} stating that for linear differential equations of the form
\begin{equation}
\frac{d}{dt}\Psi (t) = A(t) \Psi(t) \hspace{0.5cm}\text{with}\hspace{0.5cm} A(t) = A(t+T)
\end{equation}
there are periodic solutions of the form
\begin{equation}
\Psi_{\alpha} (t) = e^{-i\epsilon_{\alpha}t/\hbar} \phi_{\alpha} (t)  ,
\label{eq:Floquetsol}
\end{equation}
 where the Floquet modes have the same periodicity as the operator $A(t)$: $ \phi_{\alpha} (t) = \phi_{\alpha} (t+T)$.\cite{Floquet1883}
The Floquet Hamiltonian is defined as
\begin{equation}
\mathcal{H} (t) = H (t) - i\hbar \frac{\partial}{\partial t} ,
\label{eq:FloqH}
\end{equation}
with the Floquet modes as its eigenfunctions. This Floquet Hamiltonian can then be considered in Floquet space, a composite space of the real space $\mathcal{R}$ and the space of time periodic functions $\mathcal{T}$.
Its basis is given by 
\begin{equation}
\ket{i,k} = \ket{i} \otimes \ket{k},
\end{equation}
with $\ket{i} = c^{\dagger}_i \ket{0}$ and the Fourier basis $\ket{k}$ with $k\, \epsilon\, \mathbb{Z}$. 
The elements of the Hamiltonian in the Fourier space $\mathcal{T}$ are defined via a Fourier transformation as
\begin{equation}
\bra{k}H\ket{k'} = H_{k,k'} = \frac{1}{T} \int_0^T dt e^{i(k-k')\Omega t} H(t) ,
\end{equation}
with the driving frequency defined as the inverse of the period $\Omega = 2\pi/T$. 

With the Fourier series of the time dependent parameters 
\begin{equation} 
\epsilon_{ij} (t) = \sum_k \epsilon_{ij;k}\, e^{ik\Omega t}
\label{eq:parseries}
\end{equation}
the single particle part of the Floquet dot Hamiltonian becomes ($\hbar = 1$)
\begin{equation}
\bra{i,k}\mathcal{H}\ket{j,k'} = (\epsilon_{ij;0} - k\Omega \delta_{ij}) \delta_{kk'} +\epsilon_{ij;k'-k}  
\label{eq:e_eff_coeff}
\end{equation}
in Floquet space, where all time independent parameters are diagonal in $\mathcal{T}$; all time dependent parameters contribute with the appropriate Fourier coefficients. 

The resolvent of the Floquet Hamiltonian in Floquet space is the non-interacting, retarded Green's function
\begin{equation}
g^{\text{ret}}(\omega) = \frac{1}{\omega - \mathcal{H}+i0^+} \text{.}
\label{eq:GretFloquet}
\end{equation}

\subsection{Transformation to Floquet space} 
The starting point are the previously defined two-time Green's functions and the well known Dyson equation
\begin{equation}
\Big (i \frac{\partial}{\partial t}- \hat{\epsilon}(t)\Big ) \hat{G} (t,t') -  \int^{\infty}_{t_0} dt_1 \hat{\Sigma} (t,t_1) \hat{G} (t_1,t') = \delta (t-t') \mathbb{1},
\label{eq:Dysont}
\end{equation}
where multiplications are understood as summation over quantum numbers and $\hat{\epsilon}(t)$ is the single particle part of the Hamiltonian. 
All objects are read as matrices, ordered in the convention of Ref.\,\onlinecite{Larkin1975}
\begin{equation}
\hat{G} = 
\begin{pmatrix}
  G^{\text{ret}} & G^K  \\
  0 &  G^{\text{adv}} \\
 \end{pmatrix}.
\end{equation}
Aiming at the long time behavior, we set $t_0$ to $-\infty$, which allows to apply the following transformation to Floquet space.\cite{Tsuji2008, Rentrop2014, Genske2015}
The second time argument of the general Green's functions for $X=\text{ret, adv, K}, <,>$ is Fourier transformed,\cite{Wu2008}
 \begin{equation}
 G_{ij}^X(t,\omega)=\int_{-\infty}^{\infty} dt' e^{i\omega (t-t')}G_{ij}^X(t,t'),
 \label{eq:Greensfour1}
 \end{equation}
which in turn implies
\begin{equation}
G_{ij}^X(t,t')=\frac{1}{2\pi}\int_{-\infty}^{\infty} d\omega e^{i\omega (t'-t)}G_{ij}^X(t,\omega)  \text{.}
\label{eq:Greensfour2}
\end{equation}
Applying the Floquet theorem 
we Fourier expand with respect to the remaining time dependence
\begin{equation}
G^X_{i,j}(t,\omega)=\sum\limits_k G^X_{ij;k}(\omega) e^{-i k\Omega t},
\label{eq:Greensfourk1}
\end{equation}
with the coefficients defined as
\begin{equation}
G^X_{ij;k}(\omega) = \frac{|\Omega|}{2\pi} \int_0^T dt\, e^{i k\Omega t} G^X_{ij}(t,\omega)
\label{eq:Greensfourk2}
\end{equation}
resulting in an extra Floquet index $k$. 
The Fourier coefficients of the Green's function are understood as
\begin{equation}
G^X_{ij;k}=G^X_{ij;k0},
\end{equation}
which by using the relation
\begin{equation}
G^{\text{X}}_{ij;kk'}(\omega) = G^{\text{X}}_{ij;k-k'0}(\omega+k'\Omega) 
\label{eq:k1eff}
\end{equation}
can be generalized to
\begin{equation}
G^{X}_{ij;kk'} (\omega)= \bra{i,k}G^{X}(\omega) \ket{j,k'}
\end{equation}
in Floquet space. Equation (\ref{eq:k1eff}) introduces an artificial dependency on a second Floquet index, which allows for a simple matrix multiplication in Floquet space.

The known symmetries of the two time Green's functions and self-energies \cite{Kennes2012a} still apply, resulting in
\begin{equation}
G_{ij;kk'}^{\text{adv}} (\omega) = [G_{ji;k'k}^{\text{ret}}(\omega)]^* 
\end{equation}
and 
\begin{equation}
G_{ij;kk'}^{\text{K}} (\omega) = - [G_{ji;k'k}^{\text{K}} (\omega)]^* \text{.}
\end{equation}
The same relations hold equally for the respective self-energies.

The transformation of the Dyson equation (\ref{eq:Dysont}) yields 
\begin{equation}
\left[(\omega + k\Omega)  \dunderline{\mathbb{1}} -(\dunderline{\epsilon}+\dunderline{\Sigma}^{\rm ret}(\omega))\right] \dunderline{G}^{\rm ret} = \dunderline{\mathbb{1}}
\end{equation}
for the retarded component. The double underlines indicate the objects as matrices in Floquet space, assuming summation over the real space quantum numbers as well as the Floquet index. The identity is defined accordingly as $\dunderline{\mathbb{1}}_{ij,kk'} = \delta_{ij} \delta_{kk'}$ and $\dunderline{\epsilon}$ indicates the matrix of the single particle part of the Hamiltonian with the coefficients defined in Eq.\,(\ref{eq:e_eff_coeff}).

The structure of our approach allows to set up a flow equation, which can be equally transformed to Floquet space to compute the self-energy, which includes interaction in the system and is applicable in the whole range of driving frequency (see Sect.\ref{sec:fRG}).

\subsection{Reservoir self-energy}
The influence of the reservoirs is projected onto the dot, resulting in a reservoir self-energy $\dunderline{\Sigma}_{\rm res}$.\cite{DissCK}
It is added to the self-energy, which incorporates the interaction, when computing the full Green's function via Dyson's equation as
\begin{equation}
\dunderline{G} = (\dunderline{g}^{-1} - \dunderline{\Sigma}_{\rm res} - \dunderline{\Sigma})^{-1} = (\dunderline{G_0}^{-1} - \dunderline{\Sigma})^{-1}.
\end{equation}
In the second step the exact reservoir self-energy $\dunderline{\Sigma}_{\rm res}$ is incorporated in the effective dot Green's function $G_0$.
The Keldysh self-energy for reservoir $\alpha$ is then
\begin{align}
\Sigma_{\rm res,ij,kk'}^{\alpha, \rm{K}}(\omega) = \sum_{k_1,q_{\alpha}} v_{q_\alpha i, k-k_1}^{*}  [1-2\, f_{\alpha}(\omega+k_1\Omega)] \nonumber \\ \left[ g_{\rm res}^{\alpha, \rm ret}(\omega) - g_{\rm res}^{\alpha, \rm adv}(\omega)\right] v_{q_\alpha,j,k_1-k'},
\end{align}
where $f_{\alpha}$ is the Fermi function of reservoir $\alpha$.\\
In the case that all time dependency is on the dot, it can be simplified by an analog of the dissipation fluctuation theorem
\begin{equation}
\dunderline{\Sigma}_{\rm res}^{\alpha, \rm{K}}(\omega) = [\dunderline{\mathbb{1}}-2\, \dunderline{F}^\alpha(\omega)] \left[ \dunderline{\Sigma}_{\rm res}^{\alpha, \rm ret}(\omega) - \dunderline{\Sigma}_{\rm res}^{\alpha,\rm adv}(\omega)\right],
\end{equation}
with $\dunderline{F}^{\alpha}(\omega)_{kk'} = f_\alpha(\omega + k \Omega) \delta_{k,k'}$.

\subsection{Observables} 
The single-particle observables we are interested in can be computed from the Green's function in Floquet space. They inherit the external periodicity which allows us to write them as a sum of higher harmonics. The time dependent expectation value of the occupation number is given by  
\begin{equation}
\bar{n}_i(t)=\langle c^{\dagger}_i c_i \rangle =   \sum_k n_{i,k} e^{ik\Omega t},
\end{equation}
where
\begin{equation}
n_{i,k} = \frac{1}{4\pi i} \left[\int d\omega  G_{ii;-k0}^{\rm{K}}(\omega)\right] + \frac{1}{2} \delta_{k,0} \text{ .}
\end{equation}
The time dependent current for the reservoir $\alpha= \text{R,L}$ is \cite{Meir92}
\begin{equation}
J_{\alpha} (t) = -i \text{Tr}\, \rho_0 [H(t), N_{\alpha}(t)] = \sum_k J_{\alpha,k} e^{ik\Omega t} ,
\end{equation}
where $N_{\alpha}$ is the particle number operator of the reservoir $\alpha$. The coefficients are defined as
\begin{align}
& J_{\alpha,k}= \,\frac{1}{4\pi} \sum_{k'} \int_{\infty}^{\infty} d\omega\,\Big[\hat{\Sigma}^{\text{ret}}_{\alpha,-k-k'} (\omega+k'\Omega) \,\hat{G}^K_{k'} (\omega) 
\nonumber \\  & - \hat{G}^{\text{ret}}_{-k-k'} (\omega+ k'\Omega) \,\hat{\Sigma}^{\rm{K}}_{\alpha,k'} (\omega) \Big] +  [-k \rightarrow k]^* ,
\label{eq:cur}
\end{align}
with summation over dot indices assumed.
The pumped charge per period is defined as
\begin{equation}
Q = \frac{1}{2}\int_0^T dt\, (J_{\rm L}(t) - J_{\rm R}(t))
\end{equation}
and is connected to the mean current $ J_{k=0} = J_{{\rm L},k=0} = J_{{\rm R},k=0}$ via $ J_{k=0} = Q/T$. The latter describes the average amount of pumped charge per unit time.

\section{Functional Renormalization Group}
\label{sec:fRG}

\subsection{General idea} 
The functional renormalization group (FRG) is a method to treat many-particle problems of interacting fermions or bosons. An infinite hierarchy of differential equations for the vertex functions constitutes an exact reformulation of the many-particle problem.\cite{Metzner2012}  A flow parameter is introduced to obtain this hierarchy, which can be done in several ways, one possibility is to introduce an infrared cutoff into the bare Green's function. 
The FRG procedure then employs Wilson's idea of renormalization, where throughout the flow, contributions of all energy scales are successively summed up, regularizing the resulting expressions. 

Introducing the flow parameter $\Lambda$ into the bare Green's function ($G_0  \rightarrow G_0^{\Lambda}$), such that $G_0^{\Lambda=\infty}  =0 , G_0^{\Lambda=0} =G_0$, the one-particle irreducible $n$-particle vertex functions $\gamma_n$ depend on $\Lambda$ as well. Taking the derivative with respect to $\Lambda$ results in flow equations of the form
\begin{equation}
\frac{d}{d\Lambda} \gamma^{\Lambda}_n = \mathcal{F}(\gamma^{\Lambda}_1, \gamma^{\Lambda}_2, ... , \gamma^{\Lambda}_{n+1}, \Lambda)
\end{equation}
leading to an infinite set of coupled differential equations. 
 A complete solution of this infinite set would give the exact expressions of all $n$-particle vertex functions.

Focusing on fermionic, many-particle systems, FRG has been applied in two-, one- and zero-dimensions.\cite{Metzner2012} In the context of zero-dimensional quantum dot setups, various implementations have been realized:
whereas for equilibrium problems, Green's functions in the Matsubara formalism are derived, it is also possible to consider Keldysh Green's function with real frequencies for the steady state of non-equilibrium systems\cite{Karrasch2010, Jakobs2010} or with explicit time dependency for the transient behavior.\cite{Kennes2012a} 
It is thus a next natural step to set up the flow equations in Floquet space.

The hierarchical set of flow equations can in general not be solved and the need for truncation arises. 
The first two equations of this hierarchy are the flow equations of the single particle vertex function $\gamma_1 = - \Sigma$, i.e.\,the self-energy and of the effective two-particle interaction (two-particle vertex function). Different truncation schemes have been implemented, depending on the considered system. 
We will focus on the lowest order truncation since it has been shown that it already captures the leading low energy physics for the quantum dot system of interest.\cite{Karrasch2010}

The flow equation for the self-energy in its general form is
\begin{equation}
\partial_{\Lambda} \hat{\Sigma}^{\Lambda}(\mathbb{1'};\mathbb{1}) = - \sum_{\mathbb{22'}} \hat{S}^{\Lambda}(\mathbb{2;2'}) \gamma_2^{\Lambda} (\mathbb{1',2';1,2}) \text{.}
\end{equation}
The multi-indices $\mathbb{1}, \mathbb{1'}, \mathbb{2}, \mathbb{2'}$ include the real space quantum number $i$ as well as either Matsubara frequencies $i \omega$ for equilibrium, real frequencies $\omega$ and Keldysh contour index $p$ for a steady non-equilibrium situation or a continous time variable $t$ and Keldysh contour index $p$ in the case of explicitly time dependent non-equilibrium setups.
For time periodic systems in the long-time limit, the multi-index consists of a real space quantum number $i$, a Floquet index $k$, the continous frequency $\omega$ and the Keldysh contour index $p$. 
The so-called single-scale propagator is defined as
\begin{align}
\hat{S}^{\Lambda} (\mathbb{1;1'}) &= - \hat{G}^{\Lambda} (\mathbb{1;2})\partial_{\Lambda} [\hat{G}_0^{\Lambda}(\mathbb{2;2'})]^{-1} \hat{G}^{\Lambda}(\mathbb{2';1})  \nonumber \\
&= \partial_{\Lambda}^* \hat{G}^{\Lambda} (\mathbb{1;1'}) \text{.}
\label{eq:dfnS}
\end{align}
The truncation is realized by setting the two-particle vertex $\gamma_2$ to the antisymmetrized bare interaction $\bar{u}$
\begin{equation}
\gamma_2^{\Lambda} (\mathbb{1,2,1',2'}) = -i \bar{u}_{\mathbb{1,2,1',2'}} 
\label{eq:trunc}
\end{equation}
and all higher vertices are set to zero. This is a reasonable choice for small interactions, as $\gamma_n$ is of order $\mathcal{O}(U^n)$, where $U$ denotes the amplitude of the interaction. 
In the resulting approximation scheme the feedback of the self-energy is incorporated in the single-scale propagator, such that contributions of all order of interaction are included in the diagram. The renormalized parameters are thus correct at least to the leading order $U$, but partially capture the higher order contributions by the RG procedure.
In this way FRG is able to capture power-law behavior with $U$ dependent exponents correct to its leading order.\cite{Meden2008, Karrasch2010, Metzner2012, Kennes2012a}

The Keldysh component of the self-energy does not flow\cite{Karrasch2010, Kennes2012a} in this truncation scheme, leaving us only with one flow equation for the retarded component of the self-energy. 

\begin{figure}[t]
\includegraphics[width=0.7\columnwidth]{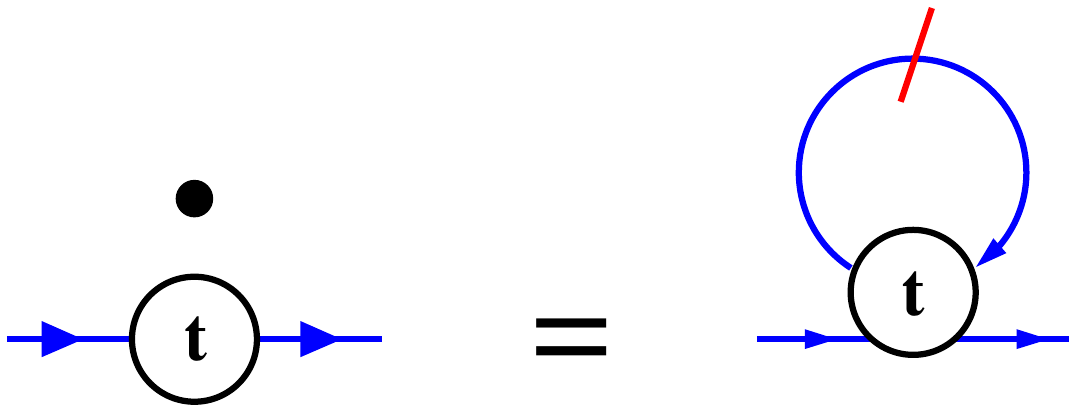}
\caption{Diagrammatic representation of the first flow equation in the time dependent form. The hierarchy of flow equations is cut after this equation by setting the two-particle vertex function to the bare interaction [see Eq.\,(\ref{eq:trunc})]. To deal with time periodic systems the flow equation is transformed to Floquet space.}
\label{fig:fRG1}
\end{figure}

Since the flow equation of the explicitly time dependent implementation of this truncation \cite{Kennes2012a} is also valid in the long time limit of time periodic setups, it is not necessary to derive the flow equations on the level of generating functionals; we can simply transform the explicit time dependent flow equation to Floquet space.  
The flow equation of the retarded component of the two-time Green's function in Keldysh space \cite{Kennes2012a} is 
\begin{equation}
\partial_{\Lambda} \Sigma_{ij}^{\text{ret},\Lambda} (t',t) = \sum_{n,m} S_{nm}^{\rm K,\Lambda} (t,t) \left[-\frac{i}{2} \bar{u}_{imjn}(t)\right]\delta(t'-t) 
\end{equation}
and is diagrammatically shown in Fig.\,\ref{fig:fRG1}.
It is transformed by Eqs.\,(\ref{eq:Greensfour1}) and (\ref{eq:Greensfourk1}), resulting in 
\begin{equation}
\partial_{\Lambda} \Sigma^{\text{ret},\Lambda}_{ij;k0} (0) = -i \sum_{n,m,k'} \int  \frac{d\omega}{4\pi} S_{nm;k'+k0}^{\rm K,\Lambda}(\omega)(\bar{u}_{imjn;k'000}) ,
\label{FRG1k}
\end{equation}
with the initial conditions
\begin{align}
\Sigma^{\rm ret, \Lambda = \infty}_{ij,k0} (0)&= \frac{1}{2} \sum_l \bar{u}_{iljl;-k000} , \label{eq:initialcond} \\
\Sigma^{\rm K, \Lambda = \infty}_{ij,k0} (0) &= 0 .
\end{align}
The right hand side of Eq.\,(\ref{FRG1k}) depends on the Keldysh Green's function which is related to the retarded self-energy via the Dyson equation. 
The lowest order truncation allows to compute the frequency independent contribution to effective, renormalized single particle parameters, rendering the system an effective non-interacting one with parameters renormalized by the interaction induced correlations at the end of the flow.

The flow equation differs by an extra single-particle-like Floquet index from the flow equation of the time independent, stationary system. These Floquet channels are coupled and higher harmonics can be created throughout the flow, rendering it necessary to solve the full flow equation numerically with an appropriate number of higher harmonics.  We will show below that in the limit of small amplitudes, the equation can be solved analytically to gain insights into the renormalization of the parameters of periodically driven quantum dot setups.

We emphasize again that no further approximation concerning either the driving frequency or the driving amplitude is necessary for the derivation of the flow equation. This must be contrasted to other approaches to treat interactions in the field of periodically driven quantum dot setups, where often the adiabatic or the antiadiabatic limit is employed to make predictions for the observables. Our only restriction is a small interaction compared to the leads bandwidth $U/D \ll 1$.

\subsection{Hybridization cut-off scheme} In order to introduce the flow parameter, we use the known reservoir cut-off scheme. \cite{ Jakobs2010, Kennes2012a} The flow parameter is the hybridization of auxiliary leads, connected to each of the dot levels. Each auxiliary reservoir is assumed to have infinite temperature, rendering its distribution function structureless and the value of its chemical potential irrelevant. Initially, these reservoirs are coupled infinitely strong, which renders all other energy scales unimportant. Throughout the flow, the auxiliary leads decouple and contributions of all energy scales are gathered to regularize possible divergencies.  At the end of the flow, the auxiliary reservoirs are completely detached, recovering the initial physical system with renormalized parameters. This  choice of the flow parameter preserves causality. \cite{DissSJ, Jakobs2010a}

\section{The Interacting resonant level model}\label{sec:model}

\begin{figure}
\includegraphics[width=\columnwidth]{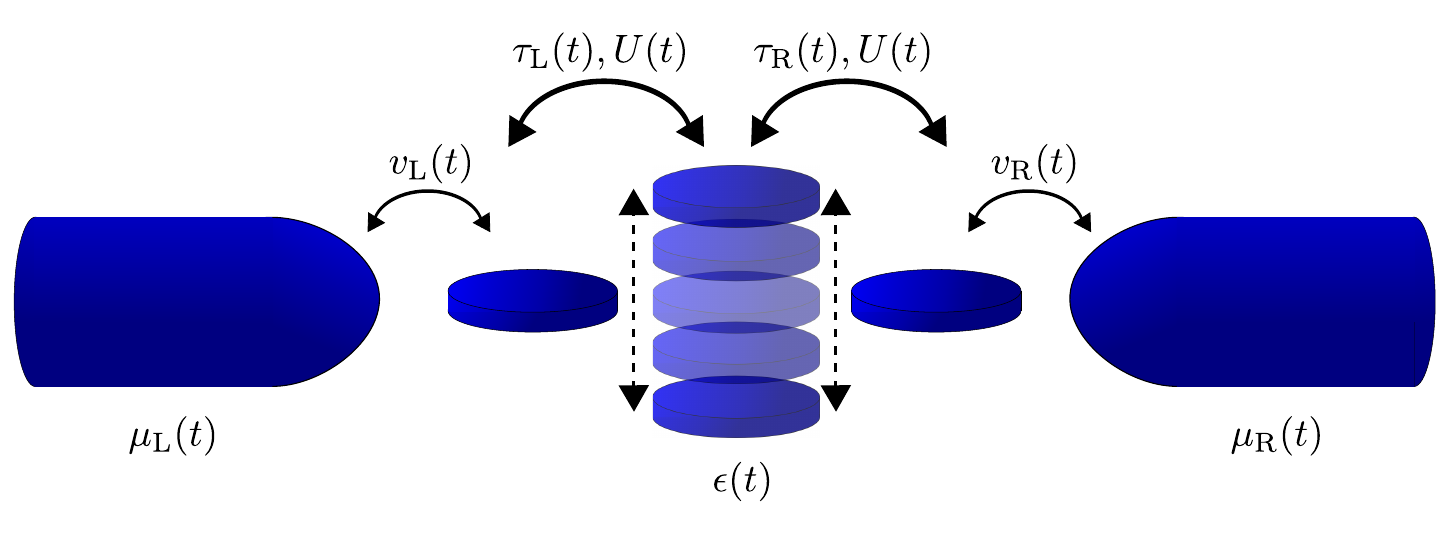}
\caption{Sketch of the interacting resonant level model, a prototype model for an interacting quantum dot. Within our method, any of the shown parameter can be considered as time periodic. The time-dependency of $\mu_{\rm L(R)} (t)$ can be shifted to the hopping amplitudes $\tau_{{\rm L(R)}}$ in the scaling limit.}
\label{fig:IRLM}
\end{figure}

\subsection{The model}
The interacting resonant level model (IRLM) has been established as a standard model  for correlated quantum dots dominated by charge fluctuations. 
It is known to be the field theoretical model of a single fermionic level coupled to two fermionic baths.
%
We realize it within a microscopic model with the dot Hamiltonian
\begin{align}
 H_{\text{dot}} =& \hspace{2mm}\epsilon (t) n_2 - \left[\tau_{\rm{L}}(t)\, d_1^{\dagger} d_2 + \tau_{\rm{R}} (t) \, d_2^{\dagger}d_3  +\text{H.c.}\right]  \nonumber \\
 +&\hspace{1mm}U \left[\left(n_1-\frac{1}{2}\right)\left(n_2-\frac{1}{2}\right) + \left(n_2-\frac{1}{2}\right)\left(n_3-\frac{1}{2}\right)\right] \nonumber \\
\end{align} 
 written in a convenient form, such that $\epsilon = 0$ corresponds to the particle-hole symmetric point. Here $n_i = d^{\dagger}_i d_i$ denotes the particle number operator of site $i$. 
The central site models the quantum dot, the first and the third site model the first side of the left and right reservoir, respectively. A Coulomb interaction $U$ between an electron on the dot and on the first site of the left and right reservoir is introduced.  We consider both cases of positive and negative interaction. A possible realization of the latter might be a quantum dot coupled to phonons, with the phonon frequency  in 
the adiabatic limit. \cite{Eidelstein2013} The resulting model is depicted in Fig.\,\ref{fig:IRLM}.

The hopping matrix elements $\tau_{\rm{L(R)}}(t)$ are time periodic with equal mean values given by the Fourier index $k=0$: $\tau_{\text{L},k=0}= \tau_{\text{R},k=0} = \tau_0$. We note that the calculations could easily be generalized to asymmetric mean values as well as different interaction strength $U_{\rm L} \neq U_{\rm R}$.

In order to reproduce the field theoretical IRLM  with this microscopic model, we employ the \textit{scaling limit}. The reservoirs are chosen to be structureless bands with hopping elements $v$ independent of the wave vector $k_{\alpha}$. The resulting bandwidth $D = \pi v^2 \rho_{\text{res}}$ with $|v| \gg |\tau_{\rm L(R)}|$ and a constant density of states $\rho_{\text{res}}$ is large compared to all other energy scales in the setup. As a consequence, a single site quantum dot is modeled with an effective hybridization of 
\begin{equation}
\Gamma_{1d} = \frac{|\tau_0|^2}{D}.
\label{eq:defG1d}
\end{equation}
Setting $D \rightarrow \infty$ and $\tau_0\rightarrow \infty$ with $\frac{|\tau_0|^2}{D} $ remaining constant, would reproduce the field theoretical model exactly. Since our lattice model interaction $U =\hat{u} (D\pi)$, where $\hat{u}$ is the interaction defined in the field theoretical IRLM, also $U/D$ is taken as a constant in this limit. \cite{Karrasch2010b}

A time dependence of the chemical potential can be shifted to the hopping amplitudes $\tau_{{\rm L(R)}}$ in the scaling limit. This is realized by a gauge transformation as e.g.\,explained in Ref.\,\onlinecite{Kaminski2000, Strass2005, Kwapinski2010, Kennes2012b}
\begin{equation}
\tau_0 \rightarrow \tau_0 e^{i\int^t_0 \mu(t') dt'} .
\label{eq:gaugetrafo}
\end{equation} 
Then the reservoir self-energies in the scaling limit are
\begin{align}
\Sigma^{\rm ret/adv}_{ij,kk'} (\omega) &= \mp i D \delta_{kk'} \delta_{ij}(\delta_{i1}+\delta_{i3}),\\
\Sigma^{\rm K}_{ij,kk'} (\omega) &= -2iD [1-2f(\omega + k\Omega)] \delta_{kk'} \delta_{ij}(\delta_{i1}+\delta_{i3}),
\end{align}
with  $D$ defined above and $f(\omega)$ is the Fermi function.

The initial condition of the flow Eq.\,(\ref{FRG1k}) for the particle-hole symmetric Hamiltonian is \begin{equation}
\Sigma^{\rm ret, \Lambda = \infty} = 0
\end{equation}
as the $U$ dependent contribution to the onsite energy cancels the initial condition Eq.\,(\ref{eq:initialcond}).

\subsection{The equilibrium IRLM} \label{sec:equIRLM}
In an equilibrium setup the self-energy behaves as $\Sigma \sim U \ln{\frac{\tau_0}{D}}$ in first order perturbation theory. It thus shows a logarithmic divergency in the wide band limit of $D \rightarrow \infty$. Several methods have been deviced to resum this divergency.\cite{Schlottmann1980a, Schlottmann1982a, Doyon2007, Borda2007, Boulat2008}  The FRG is one of these methods and has been applied succesfully both in equilibrium and non-equilibrium.\cite{Karrasch2010, Karrasch2010b}

Truncated to the lowest order an approximate self-energy is obtained within the FRG. At the end of the flow it provides correlation-induced corrections to the single-particle parameters of the Hamiltonian: the renormalized hoppings become $\tau_{\rm L}^{\rm ren} = \tau_{0}^{\rm init} + \Sigma^{\rm ret, \Lambda=0}_{12}$  as well as  $\tau_{\rm R}^{\rm ren} = \tau_{0}^{\rm init} + \Sigma^{\rm ret, \Lambda=0}_{23}$ and the renormalized onsite energy is $\epsilon^{\rm ren} = \epsilon^{\rm init} + \Sigma^{\rm ret, \Lambda=0}_{22}$. Superscript 'init' or 'ren' mark the parameter at the beginning ($\Lambda= \infty$) or at the end of the flow ($\Lambda = 0$), respectively. The interaction is thus completely reflected in the renormalization of the parameters, such that at the end of the flow the system is effectively non-interacting. 
Single-particle observables of interest for the interacting setup can be computed with the non-interacting expressions using the renormalized parameters. 

The renormalization flow of the hopping matrix element is characterized by an energy scale that provides the infrared cutoff.
An infrared cutoff is defined as the energy scale which stops the renormalization flow, i.e.\,when the flow parameter $\Lambda$ reaches the value of the infrared cutoff, the renormalization group flow levels off and saturates to its final value (compare first panel of Fig.\,\ref{fig:flussminuboth} for the equilibrium setup). The latter in turn depends in a power-law fashion on the infrared cutoff scale.
If the setup of interest features more than one energy scale, the infrared cutoff is affected by all these energy scales, which compete with each other. In the limit of one much larger energy scale compared to the others, the largest one provides the infrared cutoff.

In an equilibrium setup $\tau_0$ cuts its own flow, resulting in \cite{Karrasch2010}
\begin{align}
\frac{\tau_0^{\rm ren}}{\tau^{\rm init}_0} = \left( \frac{2 \left(\tau_0^{\rm init}\right)^2}{D^2}\right) ^{-\frac{U}{\pi D}+ \mathcal{O}(U^2)}  &\text{for}\, |\epsilon| \ll T_{\rm K}\ll D. \\
\end{align}
The emergent low energy scale $T_{\rm K}$ is defined via the charge susceptibility
\begin{equation}
\chi=\left.\frac{dn}{d\epsilon}\right|_{\epsilon=0}\sim \left(\frac{\tau_0^{\rm init}}{D}\right)^{\frac{2U}{\pi D} +\mathcal{O}(U^2)} = \frac{-\pi}{2 T_{\rm K}}.
\end{equation}
$T_{\rm K}$ will be used as the relevant low energy scale from now on.
In the non-interacting case this equals
\begin{equation}
\tilde{T}_{\rm K} = \frac{4|\tau^{\rm init}_0|^2}{D} ,
\label{eq:Tk2}
\end{equation}
where $\tau_0^{\rm init}$ can be substituted by $\tau_0^{\rm ren}$ to incorporate the interaction.
Both definitions of $T_{\rm K}$ are equivalent to the leading order in $U$ and would be the same in the limit $D \rightarrow \infty$. We thus do not differentiate between these definitions here and suppress the tilde in the following.

In the regime of $\epsilon \ll T_{\rm K}$ considered here the renormalization of the onsite energy $\epsilon$ is of the order $U^2$ (if $\epsilon^{\rm init} = 0$, $\epsilon^{\rm ren} = 0$ to all orders). Equally the onsite energies of site 1 and 3 are only renormalized to higher order.
For the role of $\epsilon$ as an infrared cutoff see Ref.\,\onlinecite{Karrasch2010}. 

\subsection{The time independent non-equilibrium IRLM} \label{sec:nonequIRLM}
 Considering a steady-state non-equilibrium setup with an applied bias voltage $V$ one finds \cite{Karrasch2010}
\begin{align}
\frac{\tau_0^{\rm ren}}{\tau^{\rm init}_0} &\sim \left( \frac{\tau_0^{\rm init}}{D}\right) ^{-\frac{2U}{\pi D}+ \mathcal{O}(U^2)}  &\text{for}\, V,|\epsilon| \ll T_{\rm K}\ll D, \label{eq:powerlaw1}\\
\frac{\tau_{0}^{\rm ren}}{\tau^{\rm init}_0}  &\sim \left( \frac{V}{D}\right) ^{-\frac{U}{\pi D}+ \mathcal{O}(U^2)} &\text{for }\, T_{\rm K},|\epsilon| \ll V \ll D  \text{ .}
\label{eq:powerlaw2}
\end{align}
The renormalization of the time independent hopping is thus characterized by the largest energy scale of the two competing energies $T_{\rm K}$ and $V$. 

For $V \gg T_{\rm K}$, $V$ cuts off the flow of $\tau_0^{\rm ren}$ and Eq.\,(\ref{eq:powerlaw2}) holds.
This power law in the voltage is also reflected in the current
\begin{equation}
J \sim \left( \frac{V}{D}\right) ^{-\frac{2U}{\pi D}+ \mathcal{O}(U^2)}  .
\end{equation}

\section{The driving frequency as an infrared scale - Analytic calculations}
\label{sec:analy}
In periodically driven dot setups the driving frequency $\Omega$ introduces a new energy scale.
To investigate its role in the renormalization flows of the parameters of the IRLM, we will consider four different protocols:
\begin{itemize}
\item In protocol 1 only the left hopping $\tau_{\rm L} (t)$ is chosen to be time periodic, while the right hopping $\tau_{\rm R}$ and the onsite energy $\epsilon$ are assumed to be time independent, i.e.\ $\tau_{\rm R} = \tau_0$ and $\epsilon =0$.
\item In protocol 2 the left and the right hopping $\tau_{\rm L(R)} (t)$ are chosen to be time periodic, while the onsite energy $\epsilon$ is assumed to be time independent and $\epsilon = 0$.
\item In protocol 3 the left and the right hopping are assumed to be time independent, i.e.\,$\tau_{\rm R} = \tau_{\rm L} = \tau_0$, while the onsite energy $\epsilon (t)$ is chosen to be time periodic. 
\item In protocol 4 the left hopping $\tau_{\rm L} (t)$ and the onsite energy $\epsilon (t)$ are assumed to be time periodic, while the right hopping is assumed to be time independent with $\tau_{\rm R} = \tau_0$. 
\end{itemize}
 As mentioned before, we concentrate on left right symmetric mean hoppings $\tau_{\text{L},k=0} = \tau_{\text{L},k=0} = \tau_0$ and the particle hole symmetric point $\epsilon_{k=0} = 0$.
The flow equations of the hopping and the onsite energy are
\begin{align}
\partial_{\Lambda} \tau^{\Lambda}_{\text{L(R)},k} &= - \frac{U}{4\pi i}\, \partial^{*}_{\Lambda} \int d\omega\, G^{\rm K,\Lambda}_{12(23);0k} (\omega), \label{eq:flowtau}\\
\partial_{\Lambda} \epsilon^{\Lambda}_{k} &= - \frac{Ui}{4\pi}\, \partial^{*}_{\Lambda} \int d\omega\, \left( G^{\rm K,\Lambda}_{11;0k} (\omega) + G^{\rm K,\Lambda}_{33;0k} (\omega) \right) \label{eq:floweps}
\end{align}
 for time independent interaction $U$ with 
\begin{align}
\tau_{{\rm L(R)},k}^{\Lambda \rightarrow \infty} = \tau^{\rm init}_{{\rm L(R)},k} , \hspace{1cm}
\epsilon_{k}^{\Lambda \rightarrow \infty} = \epsilon^{\rm init}_{k} .
\end{align}

The resulting approximate self-energies provide the correlation induced corrections to the $k$th component of the initial single particle parameters, such that the renormalized hoppings become $\tau_{\text{L},k}^{\rm ren} = \tau_{\text{L},k}^{\rm init}  + \Sigma^{\rm ret, \Lambda=0}_{12,0k}$ as well as $\tau_{\text{R},k}^{\rm ren} = \tau_{\text{R},k}^{\rm init} + \Sigma^{\rm ret, \Lambda=0}_{23,0k}$ and the renormalized onsite energy becomes $\epsilon_k^{\rm ren} = \epsilon^{\rm init} _k + \Sigma^{\rm ret, \Lambda=0}_{22,0k}$.

The four protocols are analysed analytically in the limit of small amplitudes $\Delta \tau, \Delta \epsilon$, where a dimensionless parameter $ p=\frac{\tau_{k \neq 0}}{\tau_0} = \frac{\epsilon_{k \neq 0}}{T_{\rm K}}$ is defined and kept small ($p \ll 1$). Besides of the numerical solution of the full flow equation, we aim at an analytic expression of the renormalization to leading order of $U,\frac{1}{D}, p$, but do not make any assumptions on the size of the driving frequency $\Omega$.
We reemphasize that due to the truncation the full flow equation (\ref{eq:flowtau}) and (\ref{eq:floweps}) do not contain all terms to order $U^2$ and higher. The RG procedure, however, implies a partial resummation which e.g.\,leads to power laws with $U$ dependent exponents.  In order to tackle the flow equation analytically, the respective entries of the Keldysh Green's function $G^{\rm K,\Lambda}_{ij;0k}$ are computed to order $p$ and $\frac{1}{D}$ and inserted in the flow equations (\ref{eq:flowtau}) and (\ref{eq:floweps}). The left-right symmetry of the mean hoppings as well as the interaction simplifies the analytic expressions, but analogous consideration can be made for more general expression of asymmetric setups.
For the sake of clarity, we will only present the key results of our calculations in the main text and refer the interested reader to the appendix for the details of the calculations.  
All analytic results (indicated by symbols in the figures) are compared to the full numerical solution (solid lines in the figures) of the flow equations in the following section in order to validate them.

\subsection{$k=0$ component} 

\begin{figure}
\includegraphics[width=\columnwidth]{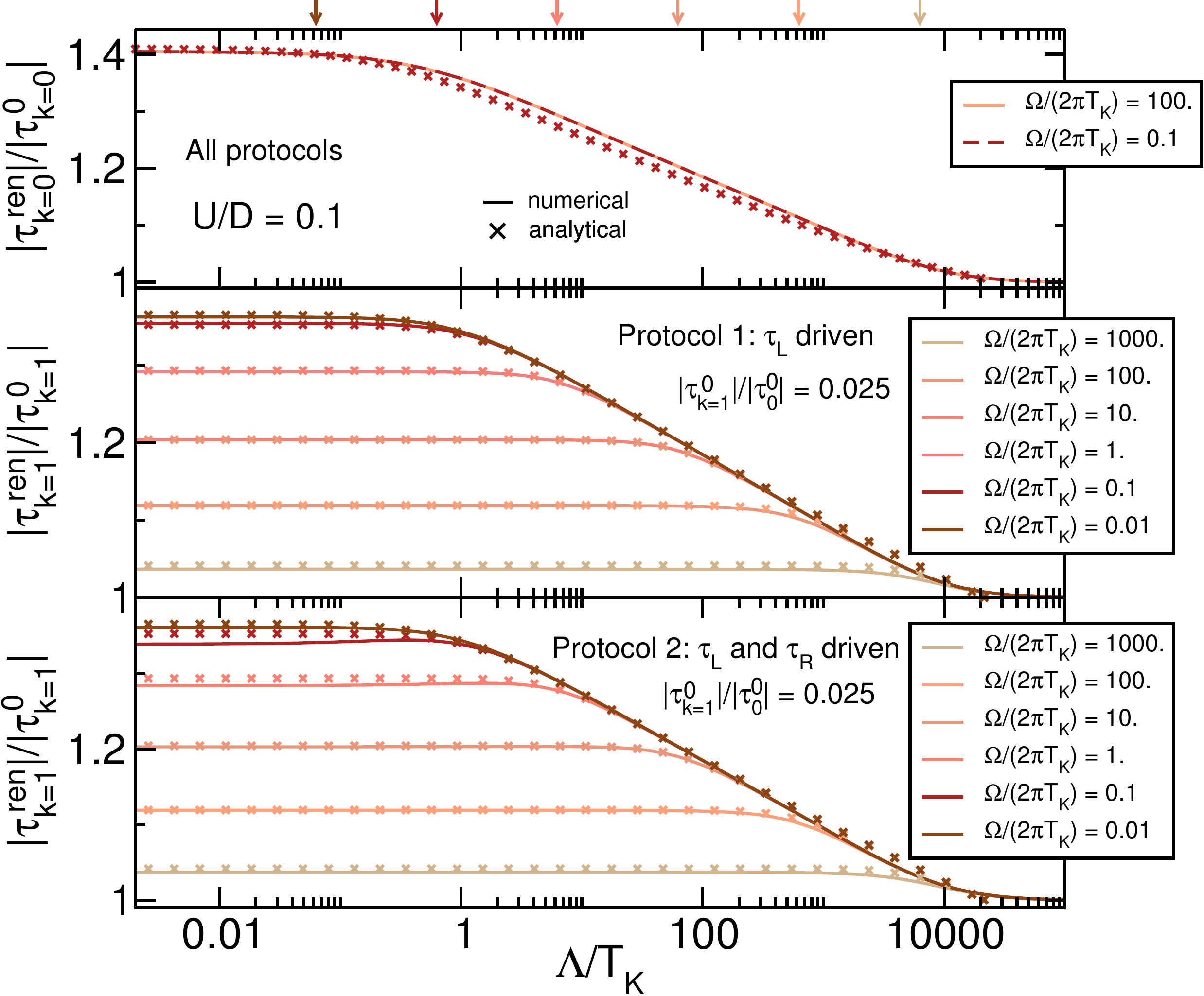}
\caption{Numerical data (solid line) and analytic expressions (crosses) for the renormalization flows of the $k=0,1$ components for protocol 1 and 2 for $U/D = 0.1$ and $T_{\rm K}/D = 4.62 \cdot 10^{-5} $ for several values of driving frequency $\Omega$. The arrows in the upper row indicate the position of the respective driving frequencies. 
The upper panel shows the $k=0$ component of the left hopping $\tau_{\rm L}$ for all protocols for different values of driving frequency. There is no dependence on the driving frequency $\Omega$. In contrast, the middle panel shows the $k=1$ component, which clearly exhibits a dependence on the driving frequency. $\Omega$ provides the cutoff of the flow if $\Omega > T_{\rm K}$ in protocol 2 if $\tau_{\rm L}$ is time periodic.  The third panel shows the flow of $\tau_{\text{L},k=1} $ if both $\tau_{\rm L}$ and $\tau_{\rm R}$ are driven. To the order $\mathcal{O}(p)$ the flows equal the ones of protocol 1, because the feedback of $\tau_{\rm R}$ is of higher order. 
}
\label{fig:flussminuboth}
\end{figure}

The analytic calculation in the limit of small amplitudes shows that the $k=0$ channel of the Keldysh Green's function decouples to the order $p$ from the higher harmonics and is independent of the driving frequency $\Omega$. As a result, the renormalization of the $k=0$ component of the left hopping $\tau_{L,0}$ can be computed independently and is unaffected by the time dependency or the exact driving setup, i.e.\,it is the same for all four protocols. 
The resulting differential equation for the mean value of the left and right hopping matrix element reproduces the one of the equilibrium setup \cite{Karrasch2010}
 \begin{equation}
\partial_{\Lambda} \tau^{\Lambda}_{0}= - \frac{U}{\pi D} \frac{\tau^{\Lambda}_{0}/D}{(\Lambda/D)^2 + \Lambda/D + 2(\tau_{0}/D)^2}.
 \end{equation}
 Solving this differential equation thus results in the same power law as discussed in the equilibrium case 
 \begin{equation}
\frac{\tau_0^{\rm ren}}{\tau_0}  \sim   \left(\frac{ \tau_0}{D}\right)^{-2\alpha_{k=0}}, \;\; \alpha_{k=0} =  \frac{U}{\pi D} +\mathcal{O}(U^2) . \label{eq:threnk0} \\
 \end{equation}
Note the suppressed superscript 'init' compared to Eq.\,(\ref{eq:powerlaw1}) for a better readibility. $\tau_0$ is assumed to be the initial value from now on, if not stated otherwise.

As depicted in the upper panel of Fig.\,\ref{fig:flussminuboth} the flow is always cut by the low energy scale $T_{\rm K}$ independent of the applied driving frequency and in agreement with the discussion in the time independent IRLM.
The analytic result captures nicely the full numerical solution, where the differences results from higher order effects in $U,p,\frac{1}{D}$, which are beyond the scope of our analytics. 
In Fig.\,\ref{fig:exponenten} the exponent $\alpha_{k=0}$ of the power law of the hopping [see Eq.\,(\ref{eq:threnk0})] is displayed. It has been extracted from the full numerical solution. This is realized via a logarithmic derivative as $d\ln(\tau^{\rm ren}_{\text{L},0})/d\ln(\tau_0)$, implemented as centered differences, which is a very sensitive measure. The resulting exponent is in good agreement with the analytic prediction in the regime of small interactions. The deviation in the regime of larger interaction results from $U^2$ contributions.
From our discussion about the time independent IRLM, we know that the onsite energy does not flow away from the particle hole symmetric point, which thus holds here equally.

Let us highlight that the driving frequency $\Omega$ does not provide an infrared cutoff for the $k=0$ component of the hopping, since the effective energy scale is $k \Omega = 0$. The component is even completely unaffected by any time periodicity in the limit of small driving amplitudes.

\subsection{Protocol 1: Time periodic $\tau_L(t)$}
\label{sec:prot1}

\begin{figure}[t]
\includegraphics[width=\columnwidth]{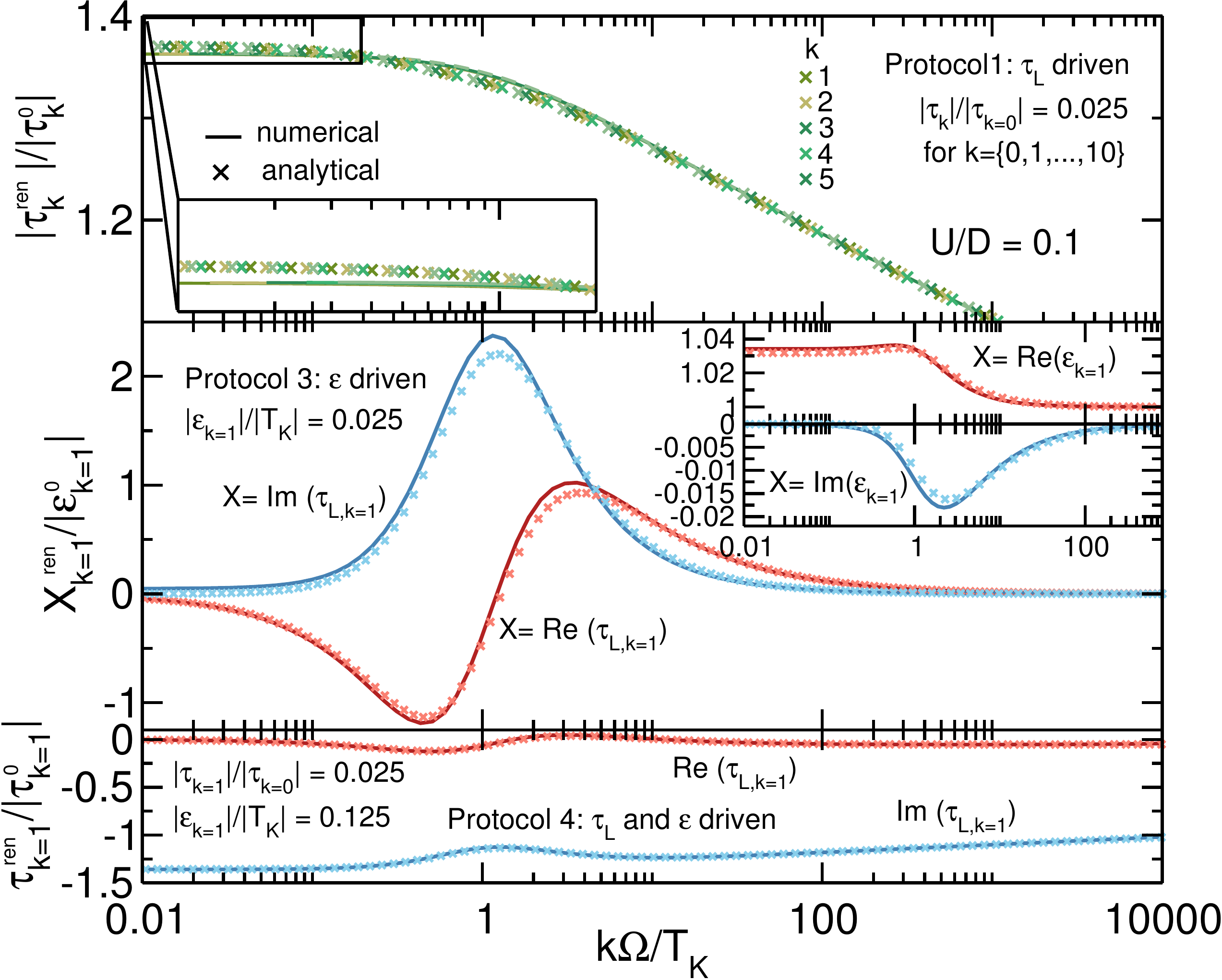}
\caption{Numerical data (sold line) and analytic expressions (symbols) for the renormalized left hopping and the onsite energy as a function of driving frequency $\Omega$ for protocols 1, 3 and 4 for $U/D = 0.1$ and $T_{\rm K}/D = 4.62 \cdot 10^{-5} $. The absolute values of the first five higher harmonics are shown in the first panel as a function of $k\Omega$. The left hopping is driven with  $\tau_{\rm L} = \tau_0 + \sum_{k=1}^{10} \Delta \tau \sin (k\Omega t)$, i.e.\,each harmonic has the same initial value. All solid lines lie on top of each other demonstrating the same functional dependence on $k\Omega$ for each $k$th harmonic. 
When $k\Omega \leq T_{\rm K}$, (shown in the inset), $\tau_{\text{L},k=1}$ bends to an $\Omega$ independent value, confirming that $T_{\rm K}$ is the main energy scale in the adiabatic limit. 
The middle panel shows the renormalization of $\tau_{\text{L}, k=1}$ as well as $\epsilon_{k=1}$ for oscillating onsite energy. While $\tau_{\text{L},k=1}$ is renormalized strongly, $\epsilon_{k=1}$ is not and its feedback into $\tau_{k=1}^{\rm ren}$ is negligible.
The third panel shows the renormalization of $\tau_{L,k=1}$ in protocol 4. Here the $\epsilon_{k=1}$ is chosen five times larger compared to the other setups, to make its contribution in form of the bump more visible.}
\label{fig:yesplusn}
\end{figure}

In protocol 1 the left hopping is varied time periodically with an arbitrary signal shape. We focus on the renormalization of the higher harmonics of the time dependent left hopping $\tau_{{\rm L},k \neq 0}$. Our analytical calculation shows (see the appendix) that the respective components of the Keldysh Green's function decouple from each other in the leading order in $p$. Moreover they show the same functional dependence on $k\Omega$, such that the $k$th coefficient only depends on the driving frequency in combination with the respective factor $k$.
As a result all higher harmonics are described by the same expression.
This allows us to obtain the following analytic expression for the flow of all $k \neq 0$ harmonics of the time periodically driven left hopping
\begin{align}
\partial_{\Lambda} \tau^{\Lambda}_{\text{L},k\neq 0}= - \frac{U}{\pi D} \frac{ \tau^{\Lambda}_{\text{L},k} \Lambda/D^2}{\frac{\Lambda^2}{D^2}\!+\!\left(\!\frac{4|\tau_0|^2}{D^2}\!+\!\frac{ik\Omega}{D}\right)\!\frac{\Lambda}{D}\!+\!\frac{2i|\tau_0|^2 k \Omega}{D^3}+\frac{4|\tau_0|^4}{D^4}} .
\label{eq:threnkn0flow}
 \end{align}
Solving Eq.\,(\ref{eq:threnkn0flow}) results in the following dependencies ($k \neq 0$)
\begin{align}
\frac{\tau^{{\rm ren}}_{{\rm L},k}}{\tau_{{\rm L}, k}}&\sim \left(\frac{\tau_0}{D}\right)^{-2\alpha_{k=1} }, \; &\alpha_{k=1}=& \frac{U}{\pi D}+\mathcal{O}(U^2),&&k\Omega \ll T_{\rm K}& \label{eq:threnkn0Tk} \\
\frac{\tau^{{\rm ren}}_{{\rm L},k}}{\tau_{{\rm L}, k}}&\sim (k \Omega)^{-\alpha_{\Omega}}, \;&\alpha_{\Omega}=& \frac{U}{\pi D}+\mathcal{O}(U^2),&&k\Omega  \gg T_{\rm K}&\label{eq:threnkn0Om}
\end{align}  
revealing power-law behavior in two different regimes:  the small frequency regime of $k\Omega \ll T_{\rm K}$ and the large frequency regime with $k\Omega \gg T_{\rm K}$. 
In the small frequency regime the $\tau_{k \neq 0}$ are cut by the energy scale $T_{\rm K}$, resulting in the same power law as for $\tau_{0}$. If the driving frequency $k\Omega$ is much larger than $T_{\rm K}$, it provides the cutoff of the renormalization flow resulting in a power law in the driving frequency. 
Hence, as discussed before, both energy scales $\Omega$ and $T_{\rm K}$ affect the infrared cutoff and compete with each other, such that in the limit of one energy scale much larger than the other, the largest one provides the infrared cutoff. This is immediately reflected in the power-law scaling of $\tau^{\rm ren}_{{\rm L},k}$.

In the central panel of Fig.\,\ref{fig:flussminuboth} the analytic expression Eq.\,(\ref{eq:threnkn0flow}) (symbols) is compared to the full numerical solution (solid line) for $\tau_{{\rm L},k=1}$ and $\tau_{{\rm L}} (t) = \tau_0 + \Delta \tau \sin (\Omega t)$. The figure shows that the analytic expression can completely capture the full numerical solution and the role of $k\Omega$ as an infrared cutoff, such that the renormalization flow bends at the respective value $\Lambda \approx k\Omega$.
The analytic description is enhanced by the feedback of $\tau_0$ by substituting the mean hopping matrix elements by their renormalized values of Eq.\,(\ref{eq:threnk0}). This has no impact in the regime of $k\Omega > T_{\rm K}$, but improves the agreement of the analytic description with the numerical solution of the full flow equation in the small frequency limit, where all components (including $\tau_0$) are cut by $T_{\rm K}$. 

\begin{figure}
\includegraphics[width=\columnwidth]{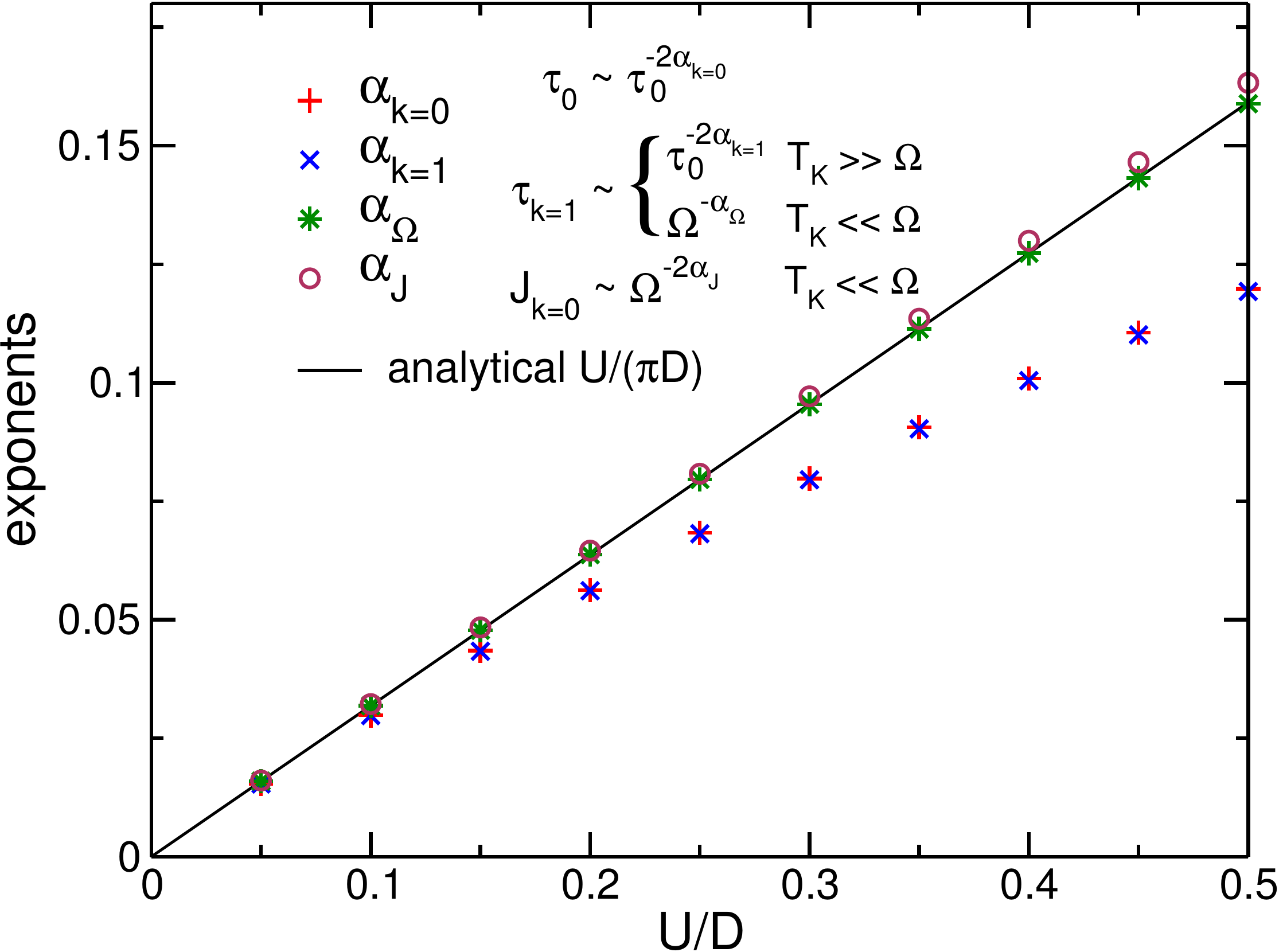}
\caption{The exponents of the renormalization of the zeroth and first harmonic of the hopping in different regimes as well as the corresponding exponent of the mean current $J_0$. All exponents are extracted from the logarithmic derivative (implemented as centered differences) of the full numerical solution. For small $U$ the exponents show excellent agreement with the analytic predictions. 
The exponents $\alpha_{k=0/1}$ show stronger higher order corrections in $U$ than the exponents $\alpha_{\Omega / J}$. }
\label{fig:exponenten}
\end{figure}

In Fig.\,\ref{fig:exponenten} the exponents of the power laws $\tau^{\rm ren}(\tau_0)$ and $\tau^{\rm ren}(\Omega)$ are displayed, which were computed by a logarithmic derivative of the numerical solution of the full flow equation.
Both exponents $\alpha_{k=1}$ and $\alpha_{\Omega}$ show excellent agreement with the analytic prediction, where the higher order corrections with increasing interaction strength are weaker for $\alpha_{\Omega}$.
The renormalized higher harmonics at the end of the flow are depicted as a function of $k \Omega$ in the upper panel of Fig.\,\ref{fig:yesplusn}, where the analytic expression is compared to the full numerical solution. Here the left hopping is varied as $\tau_L (t) = \tau_0 + \sum_{k=1}^{10} \Delta \tau \sin (k\Omega t)$ such that all non-vanishing Fourier coefficients have the same initial value. It shows the universal dependence on $k\Omega$ for each $k$th harmonic (all five solid lines lie on top of each other). The inset shows that in the limit of small frequencies all curves bend to an $\Omega$ independent value due to the infrared scale $T_{\rm K}$.

The dependency  of the renormalization of the $k$th Fourier coefficient on $k \Omega$ has interesting implications for the renormalized signal $\tau_{\rm L}^{\rm ren} (t)$: First, due to the $k$ dependency, each component is renormalized differently, i.e.\,with increasing Fourier index, the renormalization is weakened. If the driving frequency is fixed, the different strength of renormalization changes the ratio between the Fourier coefficients and thus modifies the signal shape. 
This can be utilized to design the initial signal such that the renormalized one has a desired form.\cite{Eissing16}

On the other hand, the renormalization depends on the driving frequency, such that signals with a larger driving frequency are renormalized weaker than adiabatically driven ones. 
While a positive sign of the interaction decreases the amplitude with increasing $\Omega$, the opposite is true for a negative interaction.
Thus, depending on the sign of the interaction, a rectification or amplification of the effective signal amplitude is observed.

Finally, we mention that the discussed power law in the regime of larger driving is also reflected in an observables such as the mean current $J_0$ of a single parameter pump. We postpone the detailed discussion to section \ref{sec:singleparpump}, where results for different kinds of pumping setups are presented.

\subsection{Protocol 2: Time periodic $\tau_L (t)$ and $\tau_R (t)$} 
The analytic calculation of the Keldysh Green's function shows that there is no contribution of the higher harmonics of the right hopping to renormalization of the higher harmonics of the left hopping and vice versa to the order $\mathcal{O}(\frac{1}{D}, p, U)$.
As a consequence, the flow of the left and right hoppings $\tau_{\text{L(R)},k}$ are described by the same analytic expression as for $\tau_{{\rm L}, k}$ in protocol 1. This is depicted for $\tau_{\text{L},k=1}$ in the lowest panel of Fig.\,\ref{fig:flussminuboth}. The Eqs.\,(\ref{eq:threnkn0flow}), (\ref{eq:threnkn0Tk}) and (\ref{eq:threnkn0Om}) hold equally in this setup.

\subsection{Protocol 3: Time periodic $\epsilon (t)$} 
\label{sec:prot3}

In protocol 3 only the onsite energy is driven periodically and we focus on a sinusoidal signal 
\begin{equation*}
\epsilon(t)=\Delta \epsilon \cos (\Omega t)
\end{equation*} 
around the particle hole symmetric point.

To obtain the analytic expression of the renormalization of $\epsilon^{\Lambda}_{k}$  $G^{\rm K}_{11(33),0k}$ is computed to leading order in $p$ indicating that the higher harmonics are decoupled as in protocol 1. For the sinusoidal signal, we can focus on $k=1$ only.
Since the first order perturbation theory contribution to $\epsilon_{k}^{\rm ren}$ is not plagued by logarithmic divergencies, the resulting expression 
\begin{widetext}
\begin{equation}
\epsilon^{\rm ren}_{k=1} (\Omega) \stackrel{D \gg \tau_k/\Omega}{=} \frac{U}{\pi D} \frac{T_{\rm K}}{2} \frac{\epsilon^{\rm init}_{k=1}}{D}  \frac{i(T_{\rm K}/2 +i \Omega)/D}{\Omega/D (T_{\rm K}+i \Omega)/D} \left[  \ln \left(\frac{T_{\rm K}^2}{T_{\rm K}^2+4\Omega^2} \right) - 2i \arctan\left( \frac{2\Omega}{T_{\rm K}}\right) \right]
\end{equation}
\end{widetext}
captures $\epsilon_{k}^{\rm ren}$ analytically when including the feedback of $\tau_0$.
This also holds for the higher harmonics of more general setups and is shown in the appendix.
The inset of the central panel of Fig.\,\ref{fig:yesplusn} shows that the analytic treatment (symbols) indeed captures the numerical solution of the full flow equation (solid line). While the imaginary part is mainly renormalized in the regime $\Omega \approx T_{\rm K}$, the real part is renormalized for $\Omega \lessapprox T_{\rm K}$; in both cases the renormalization is minor. 

The renormalization of $\tau^{\Lambda}_{{\rm L(R)},k=1}$ is more interesting: while its initial value is zero, it is finite throughout the renormalization flow. The renormalization of the left and right hopping is exactly the same, due to the left-right-symmetry of the setup. Evaluating the Keldysh Green's function $G^{\rm K}_{12(23),01}$ for the left (right) hopping, reveals that it does not depend on $\tau^{\Lambda}_{{\rm L(R)},k=1}$, i.e.\,the feedback of $\tau^{\Lambda}_{{\rm L(R)},k=1}$ into its own renormalization is of order $\mathcal{O}(U^2)$. 
As a result, the renormalization is not described by a differential equation, but can be computed in a first order perturbation theory calculation. 
It proves to be advantageous to do this in an effective model, which is realized by employing the replica idea, as e.g.\,discussed in Refs.\,\onlinecite{Shirley1965, Zeldovich1967, Gomez2013}. The time periodic system is mapped on a time independent system with an infinite number of replicas in an auxiliary $k$ direction, when interpreting the Floquet index $k$ as an extra spatial index. The various replicas are coupled by the higher harmonics of the time periodic parameters. For an effective model, only those replicas are included, which can interfere with $\tau_{{\rm L},k=1}$ to leading order of $p$. From this we can compute the renormalization of the hopping elements $\tau_{\rm L(R)}$ at the end of the flow
\begin{widetext}
 \begin{align}
\tau^{\rm ren}_{\rm{L(R)},k=1} \stackrel{D \gg \tau_k/\Omega}{=} -&\frac{U}{2i\pi D} \frac{\tau^{\rm ren}_{0}}{(T_{\rm K}+i\Omega)/D} \frac{\epsilon_{k=1}}{D} \left[ - 2 i \arctan\left(\frac{T_{\rm K}}{2\Omega}\right) + i\pi  +\ln\left(\frac{T_{\rm K}^2+4\Omega^2}{T_{\rm K}^2}\right)\right].
\end{align}
\end{widetext}
A more detailed explanation of the effective model as well as the analytic calculations is given in the appendix.

In the upper panel of Fig.\,\ref{fig:flussepsoundpump} the result of the perturbative calculation (crosses) is compared to the full numerical solution (solid line) for the renormalization flow of $\tau_{L,k=1}$. The flow diagram shows more structure compared to the afore discussed one and three regimes can be identified: For large driving frequencies the hoppings are barely renormalized, resulting in minor renormalization of $\tau_{{\rm L(R)},k=1}$. In the adiabatic regime on the contrary, the renormalization is sizable, but decreases again when the flow parameter reaches the regime $\Lambda \lesssim T_{\rm K}$. In the regime of moderate driving frequency $\Omega \approx T_{\rm K}$ (dark red line) renormalization is the strongest.
The renormalization flow thus features not only one infrared cutoff, but is affected by a complex interplay of $\Omega$, $\epsilon_{k=1}$, $T_{\rm K}$ and the flow parameter $\Lambda$.
\begin{figure}
\includegraphics[width=\columnwidth]{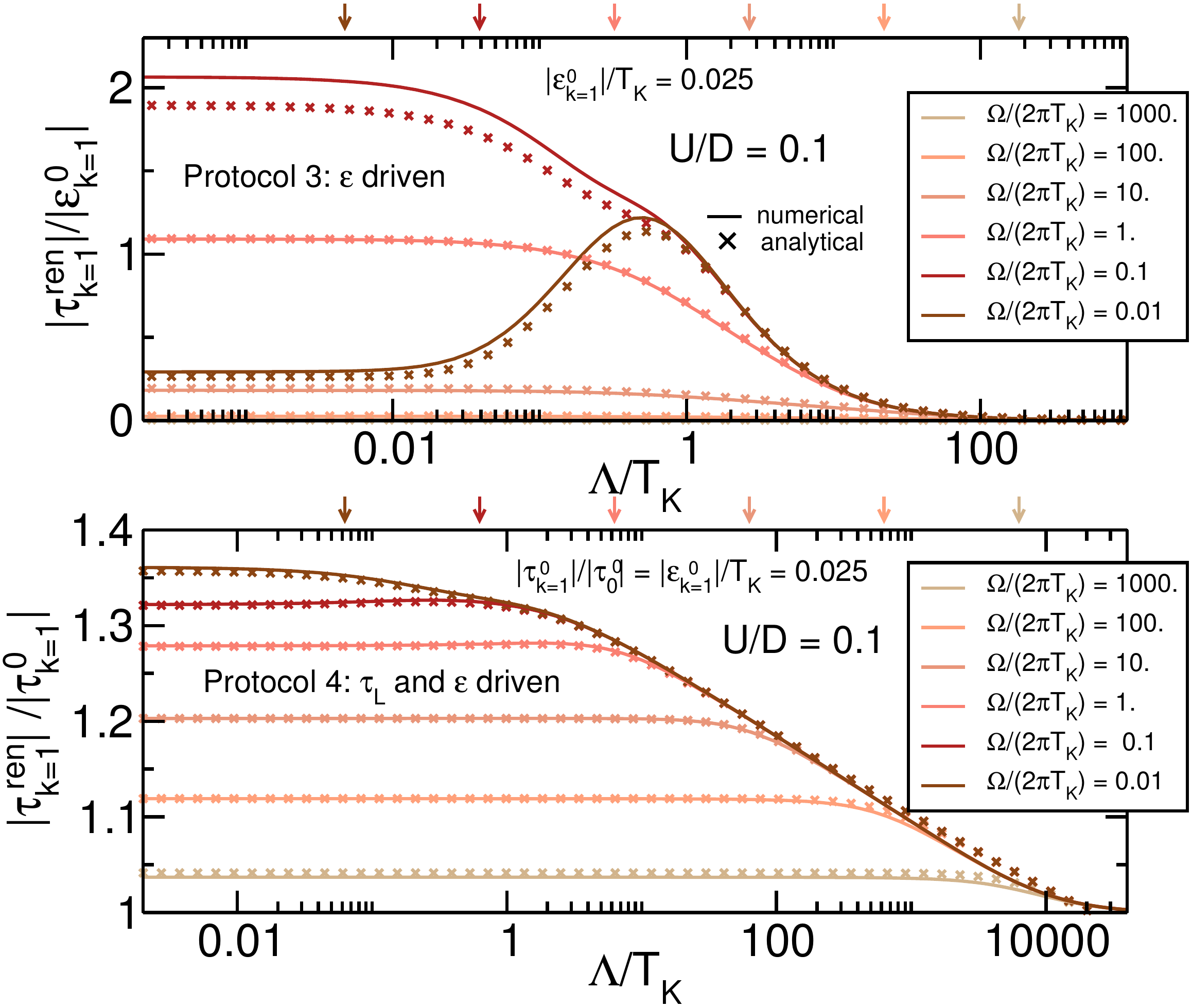}
\caption{Numerical data (solid lines) and analytic expressions (symbols) of the renormalization flows of $\tau_{k=1}$ for protocol 3 and 4 for $U/D = 0.1$ and $T_{\rm K}/D = 4.62 \cdot 10^{-5}$ for several values of the driving frequency $\Omega$. The arrows on the upper line indicate the respective value of $\Omega$.
The upper panel shows the renormalization group flow of the $k=1$ component of protocol 3, which is more complicated than the renormalization of the $k=1$ component of the protocols 1 and 2. While in the antiadiabatic limit the renormalization is surpressed by the driving frequency $\Omega$, the interplay between $\Omega$ and $\Lambda$ at around $T_{\rm K}$ results in strong renormalization for $\Omega \approx T_{\rm K}$ (compare also Fig.\,\ref{fig:yesplusn}).
The lower panel shows the renormalization of $\tau_{k=1}$ in protocol 4. The renormalization for the large values of  driving frequency $\Omega$ is only defined by the contribution of the driven $\tau_{\rm L}$, in the limit of adiabatic driving the flow shows deviations resulting from the contribution by the renormalization due to the nonzero $\epsilon_{k=1}$.}
\label{fig:flussepsoundpump}
\end{figure}
The renormalized hopping is shown in the central panel of Fig.\,\ref{fig:yesplusn}, where the analytic expression (crosses) and the full numerical solution (solid line) are presented for $\tau_{{\rm L},k=1}$. The real and imaginary parts are shown featuring the most prominent renormalization for $\Omega \approx T_{\rm K}$ and no renormalization in the adiabatic ($\Omega \rightarrow 0$) as well as the antiadabiatic ($\Omega \rightarrow \infty$) limit. The difference between the analytic expression and the numerical data is of order $\mathcal{O}(U^2)$ and thus beyond the scope of our considerations.


\subsection{Protocol 4: Time periodic $\tau_L (t)$ and $\epsilon (t)$}
In protocol 4, we assume both left hopping and onsite energy to be time periodic 
\begin{align*}
\tau_L (t) &= \tau_0 +\Delta \tau \sin(\Omega t), \\
\epsilon (t) &= \Delta \epsilon \cos (\Omega t).
\end{align*}
In this protocol, we can combine the results of the protocols 1 and 3: For the renormalization of  $\tau_{{\rm L}, k=1}$ to leading order in $U$, the two contributions of both protocols add up. As discussed in protocol 3, the feedback of $\tau_{{\rm L},k=1}$ into its own renormalization is of order $\mathcal{O}(U^2)$ and thus beyond the scope of our calculations. The renormalization of $\epsilon_{k=1}$ is the same as in protocol 3, since there is no contribution to it generated in protocol 1.

The flow of the left hopping $\tau_{{\rm L},k=1}$ is depicted in Fig.\,\ref{fig:flussepsoundpump}, comparing analytic expression (symbols) and numerical solution (solid line).  For large driving frequency the renormalization flow is mainly characterized by the contribution of protocol 1, featuring the driving frequency as the infrared cutoff. In contrast to this, the renormalization flows of $\Omega = 0.1 T_{\rm K}$ and $\Omega = 0.01 T_{\rm K}$ sheer off, reflecting pronouncedly the contribution to the renormalization induced by $\epsilon_{k=1}$.
In the lower panel of Fig.\,\ref{fig:yesplusn}, the analytic expression and the full numerical solution of the renormalized $\tau_{{\rm L}, k=1}$ is presented. In the limits $\Omega \rightarrow 0$ and $\Omega \rightarrow \infty$ the renormalization is defined by the contribution of the time periodic left hopping (from protocol 1), the contribution of the time periodic $\epsilon (t)$ manifests itself as a bump in the intermediate regime of $\Omega \approx T_{\rm K}$.

The setups of protocol 2 and 4 correspond in the adiabatic limit to the two parameter pump as described by Brouwer. \cite{Brouwer98} While for protocol 2 there is no pumped charge observed at the particle hole symmetric point, there is maximal pumping in protocol 4. We will further discuss the charged pump of this setup out of the adiabatic limit and gradually varying the phase shift between the onsite energy and the left hopping in section \ref{sec:inph}.

\section{Tuning the effective reservoir distribution function}
\label{sec:Bessel}
We leave the limit of small driving amplitude to investigate the effect of the reservoir distribution function on the renormalization flow. 
The reason we can find power laws in the limit of vanishing temperature are the sharp edges of the reservoir distribution function. They lead to divergencies in the self-energy which are summed up throughout the renormalization group flow. The respective energy scales (defining the positions of the divergencies) are reflected in the infrared cutoff, where the larger the height of the step the larger is its contribution. If the infrared cutoff is dominated by a single energy scale, we find a single dominant power law. 
Finite temperature broadens the steps and thus regularizes the divergencies in the self-energy.

In contrast to $T=0$ equilibrium where there is always a single step at the Fermi level of height one, non-equilibrium shows richer physics. The additional degree of freedom introduced by leaving equilibrium can be used to control and tune the effective reservoir distribution function. A well known example, is a symmetrically applied bias voltage $V$, where the resulting steps in the distribution function are at $\omega = \pm \frac{V}{2}$. In the case of $V \gg T_{\rm K}$ this then results in the power law discussed in Sec.\,\ref{sec:nonequIRLM}.
Nevertheless, the tunability in this setup is comparably limited, when contrasted to time periodically driven systems, which allow to tune the effective reservoir distribution function in a very diverse way creating intriguing multi-step situations. 


To make use of this we turn to the setup of protocol 3 where the onsite energy is varied 
\begin{equation}
\epsilon (t) = \Delta \epsilon \cos (\Omega t)
\label{eq:Besselsetup}
\end{equation}
around the particle-hole symmetric point. 

In order to illustrate the physical situation at hand, the time dependency of the dot is shifted to the reservoir couplings $v(t) = v_0 \exp(i\int_{t_0}^t  \Delta \epsilon \cos (\Omega t') dt')$ as described by Eq.\,(\ref{eq:gaugetrafo}) and in Refs.\,\onlinecite{Kwapinski2010, Suzuki2015}. We note that in the numerical implementation we do not perform this gauge transformation and stick to the scheme described above. The Keldysh self-energy is 
\begin{equation}
\Sigma^{\rm K}_{\alpha,kk'} (\omega) = -2iD [\delta_{kk'} -2\tilde{f}_{\alpha,kk'}(\omega)],
\end{equation}
with an effective reservoir distribution function
\begin{equation}
\tilde{f}_{\alpha,kk'}(\omega) = \sum_{k_2} J^{*}_{k-k_2}\left( \frac{\Delta \epsilon}{\Omega}\right) \hspace{0.5mm} f_{\alpha}(\omega + k_2 \Omega) \hspace{0.5mm} J_{k'-k_2}\left( \frac{\Delta \epsilon}{\Omega}\right)
\end{equation}
also referred to as the generalized distribution function.
Here $J(x)$ is the Bessel function and the bandwidth $D = \pi \rho_{\rm res} |v_0|^2$ is defined as before.
To consider the effect on the mean value of the hopping matrix elements, we focus on the $k=0$ component of $\tilde{f}_{\alpha} (\omega)$, which is given as the weighted sum of the Fermi distribution function
\begin{equation}
\tilde{f}_{\alpha, 0}(\omega) = \sum_m \left[ J_m\left(\frac{\Delta \epsilon}{\Omega}\right)\right]^2 f_{\alpha}(\omega + m\Omega),
\end{equation}
where $f_{\alpha}(\omega) = (e^{\beta_{\alpha}\omega}+1)^{-1}$. 
At $\beta_\alpha\to \infty$ it shows a multistep structure with steps of width $\Omega$ and a height determined by the $n$th Bessel function 
\begin{equation}
h_n = \left|J_n \left(\frac{\Delta \epsilon}{\Omega}\right)\right|^2
\end{equation}
 at $\omega = n\Omega$.\cite{Hettler1995, Suzuki2015, Bruder1994}
It is thus possible to tune the effective reservoir distribution by selecting a certain ratio of amplitude and driving frequency $q= \frac{\Delta \epsilon}{\Omega}$. 
We here study three different cases.

\begin{figure}
\begin{minipage}[t!]{0.74\columnwidth}
\includegraphics[width=\textwidth]{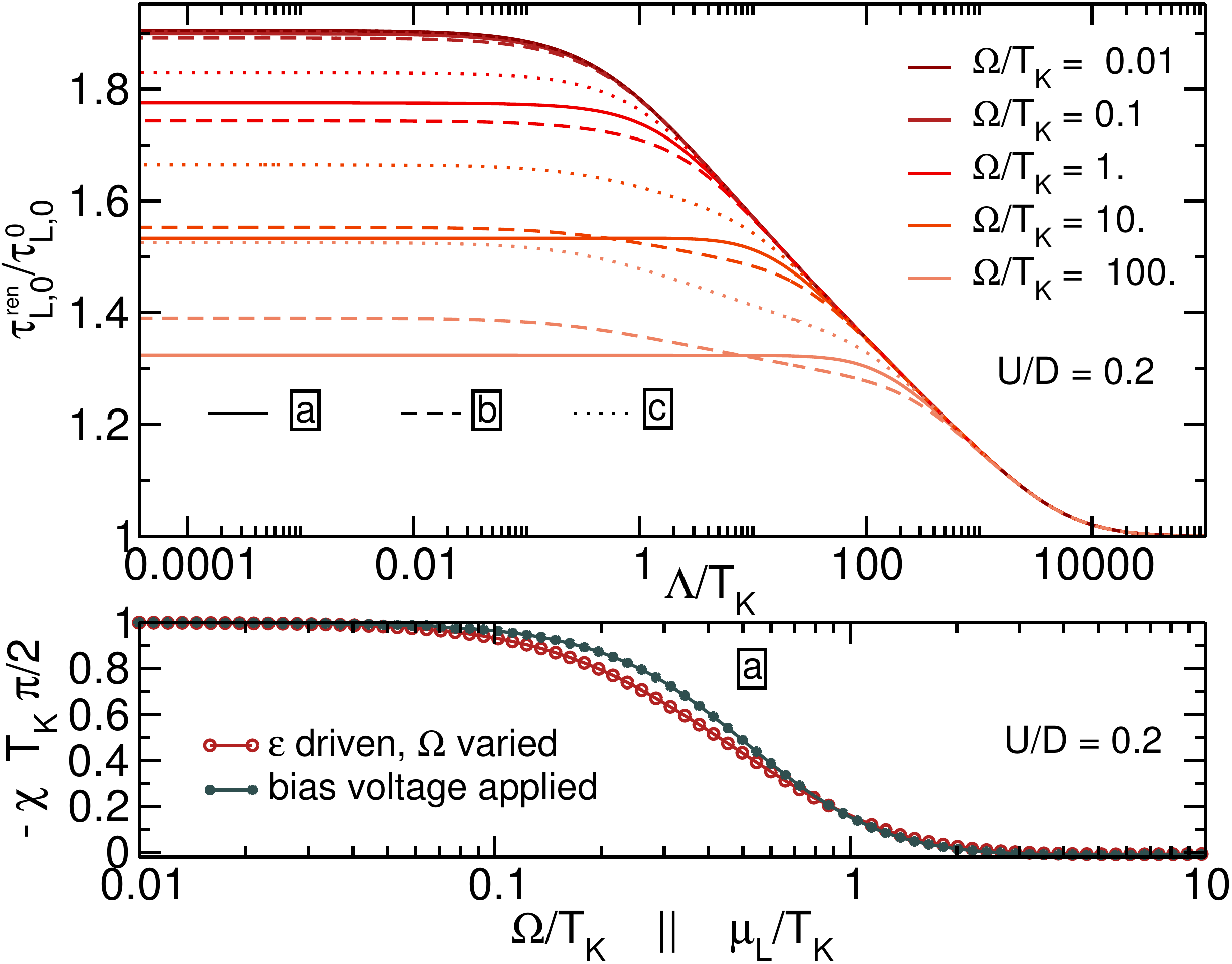} 
\end{minipage}
\begin{minipage}[t!]{0.19\columnwidth}
\vspace{1mm}
\includegraphics[width=\textwidth]{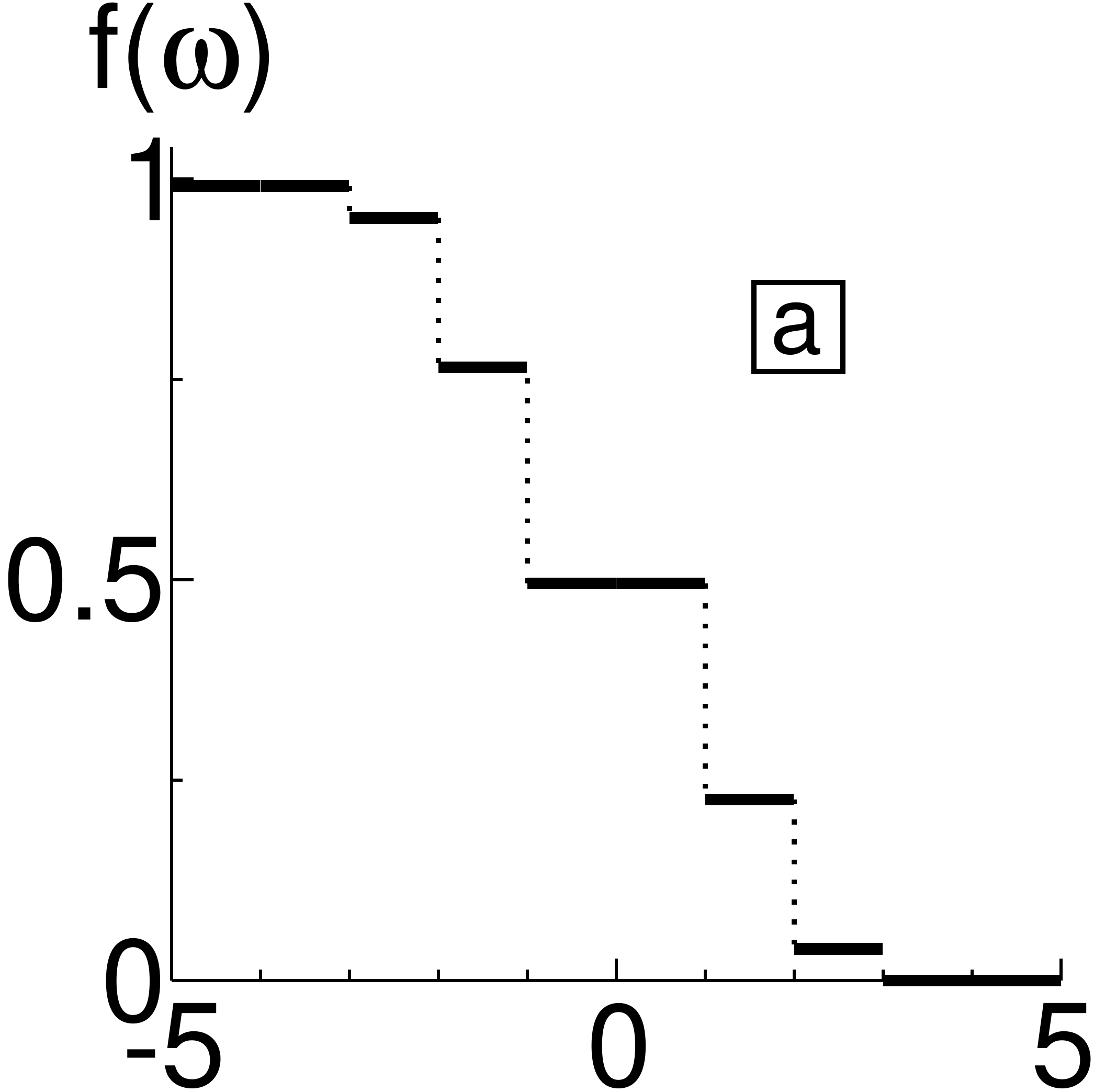}
\vspace{1mm}
\includegraphics[width=\textwidth]{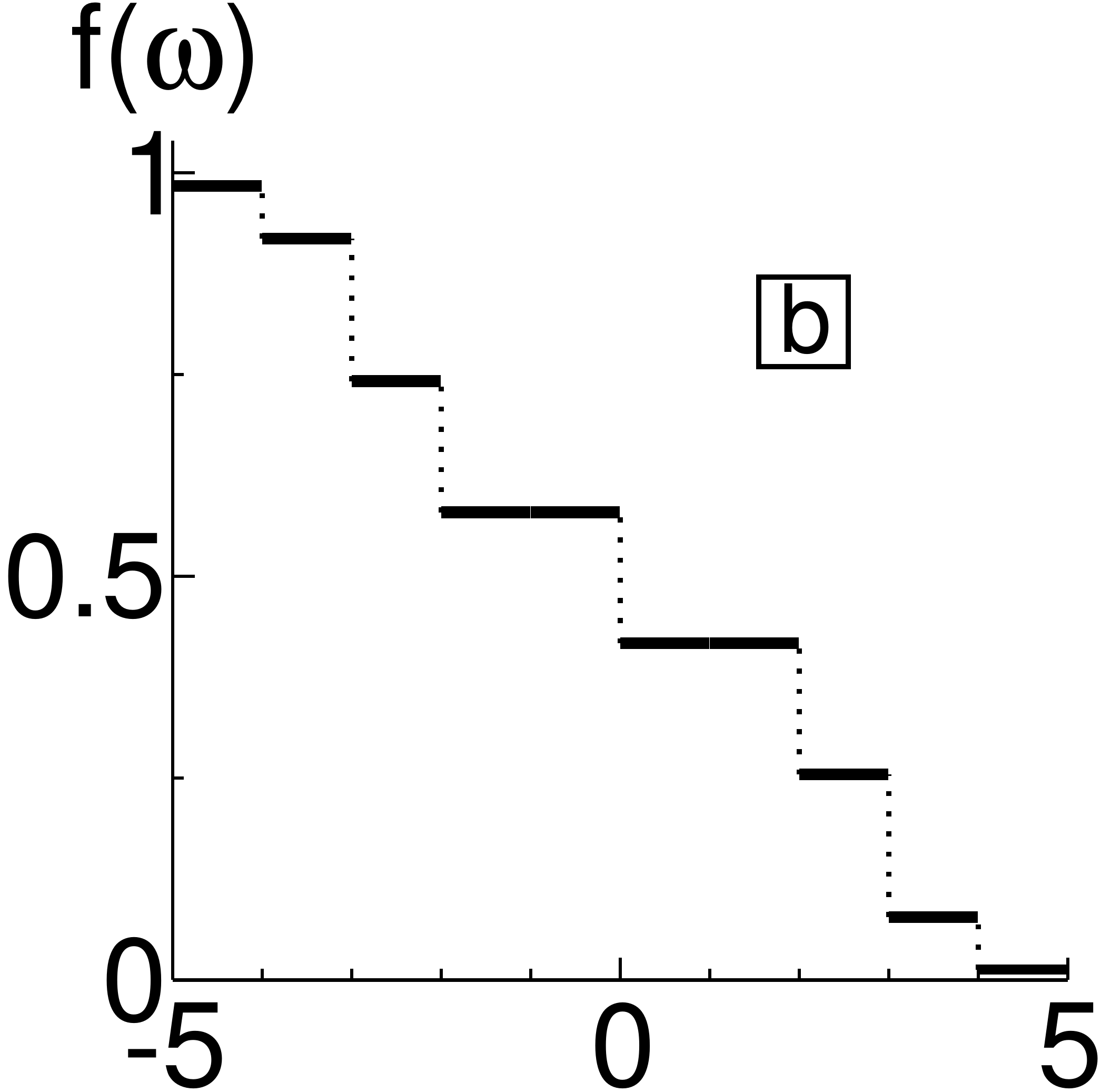}
\vspace{1mm}
\includegraphics[width=\textwidth]{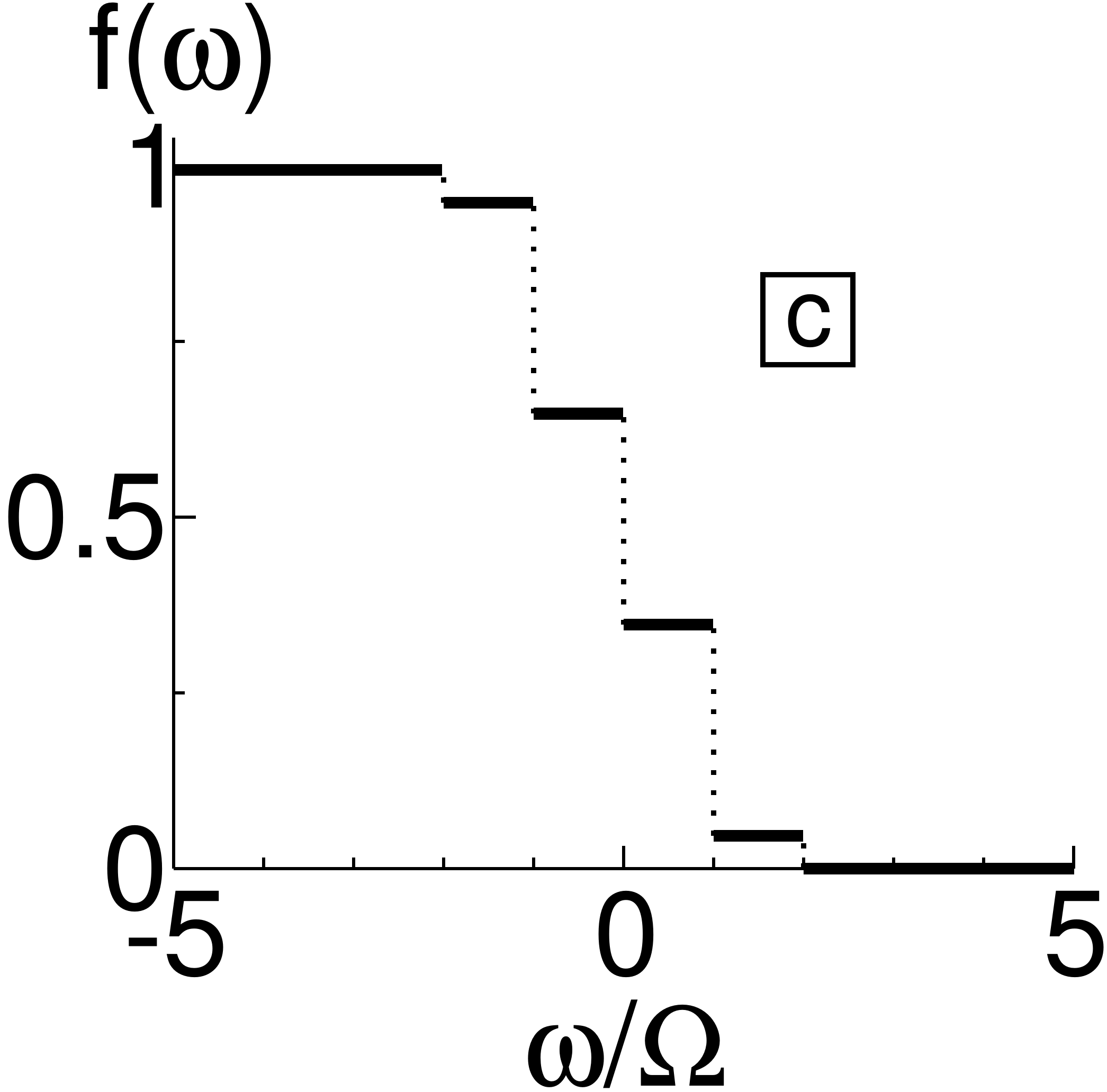}
\vspace{1mm}
\end{minipage}
\caption{The frequency $\Omega$ can act as cutoff even in the zeroth component: Driving only the onsite energy $\epsilon_0$ results in a multistep effective reservoir distribution functions, with steps at $\omega = n\Omega$  of width $\Omega$ and height  $h_n = |J_n (\frac{\Delta \epsilon}{\Omega})|^2$. The distribution functions are sketched at the right hand side to picture the different situations. The effective distribution function is then reflected in the renormalization flow of the $k=0$ harmonic of the hopping. The renormalization (solid line) in protocol (a) shows a clear infrared cutoff due to the edge at $\omega = \Omega$.  The renormalizations (dashed line) in protocols (b) shows bending at two energy scale due to the edges at $w =0, \pm 2\Omega$. The renormalization (dotted line) in protocol (c) is characterized by the multiple energy scale off the equidistant steps with equal height, resulting in a long tail.
In the lower panel, the susceptibility of protocol (a) is compared to the susceptibility of a setup with an applied DC bias voltage. Due to similar renormalization of the zeroth component of the hopping, the susceptibility shows equal behavior for both situations. For the whole plot $U/D = 0.2$ with $T_{\rm K}/D = 7.93 \cdot 10^{-5}$.}
\label{fig:Besselroot}
\end{figure}

In case (a) the ratio is fixed by $J_0(q) =0$ ($q \approx 2.405$). The resulting effective reservoir distribution of the $k=0$ channel (right top of Fig.\,\ref{fig:Besselroot}) shows no step at $\omega=0$, but steps at $\omega = \pm  \Omega$. This way we have designed an effective reservoir function that resembles in the regime of small energies $|\omega|$ the form of the effective reservoir distribution of the time independent dot model with an applied bias voltage $V = 2 \Omega$.  The renormalization group flow of the $k=0$ component of the left hopping is depicted as the solid line in Fig.\,\ref{fig:Besselroot} for several values of the driving frequency. 
The flow is clearly characterized by an infrared cutoff at $2\Omega$ as long as the driving frequency is larger than $T_{\rm K}$. This follows from the positions of the steps in the reservoir distribution function located at the driving frequency in strict analogy to a time-independent setup with a driving bias voltage. 
The resulting divergencies at $\Omega$ sum up to a cutoff in the infrared for the renormalization flow of the $k=0$ coefficient of $\tau_{\rm L}$.
The steps at larger $\omega$ do not affect the flow significantly due to their smaller heights.

In the lower panel of Fig.\,\ref{fig:Besselroot} the charge susceptibility of case (a) and of a setup with an applied bias voltage as discussed in Ref.\,\onlinecite{Karrasch2010} are compared. This observable is determined by the renormalized hoppings and thus shows equal behavior for both situations, confirming the afore discussed similarity in the  renormalization flows.

In case (b) we choose $J_1(q) =0$ ($q \approx 3.830$) for the steps at $\omega = \pm \Omega$ to vanish. The effective distribution function is shown on the right hand side of Fig.\,\ref{fig:Besselroot}. It features steps at $\omega =0$ and $\omega = \pm 2\Omega$. The flow (dashed line in Fig.\,\ref{fig:Besselroot}) exhibits a less clear saturation behavior and is cut off twice: First at the scale $2\Omega$ reflecting the edge at $\omega = 2\Omega$ and then subsequently around $T_{\rm K}$, reflecting the edge at $\omega = 0$. 

Finally, in case (c), the ratio $q$ is chosen such that $J_0 (q) = J_1 (q)$ with $q \approx 1.435$, i.e.\,with edges at $\omega = 0, \pm 1 \Omega$ with equal height. Consequentially, 
the equally separated edges all contribute likewise and are reflected as multiple energy scales in the RG flow. The resulting flow (dotted line) does not feature one pronounced infrared cutoff but is rather cut by each of the equally distant energy scales. As a consequence, the flow is bend in a long tail.

In contrast to the limit of small amplitudes (see Sec.\,\ref{sec:analy}), the $k$th component of the hopping is no longer only defined by a dependency on $k\Omega$. In particular, already $\tau_{k=0}$ depends on $\Omega$.  Since our approach follows a transparent renormalization group procedure and is not biased by any assumption made in the process of setting up the RG equations, it is fully capable to capture this dependency.  Unequivocal dependencies in the RG flow such as the $k\Omega$ dependency in the $k$th coefficient in the small amplitude limit might be captured by alternative RG methods as well. However, capturing dependencies, which result from a more involved interplay of cut-off scales, is often quite complicated in other RG approaches than the FRG. Therefore, this is an example where FRG naturally shows its full potential.

\section{Quantum pumps out of the (anti)adiabatic limit}
\label{sec:pump}
\subsection{In phase quantum pump}
\label{sec:inph}
\begin{figure}
\includegraphics[width=\columnwidth]{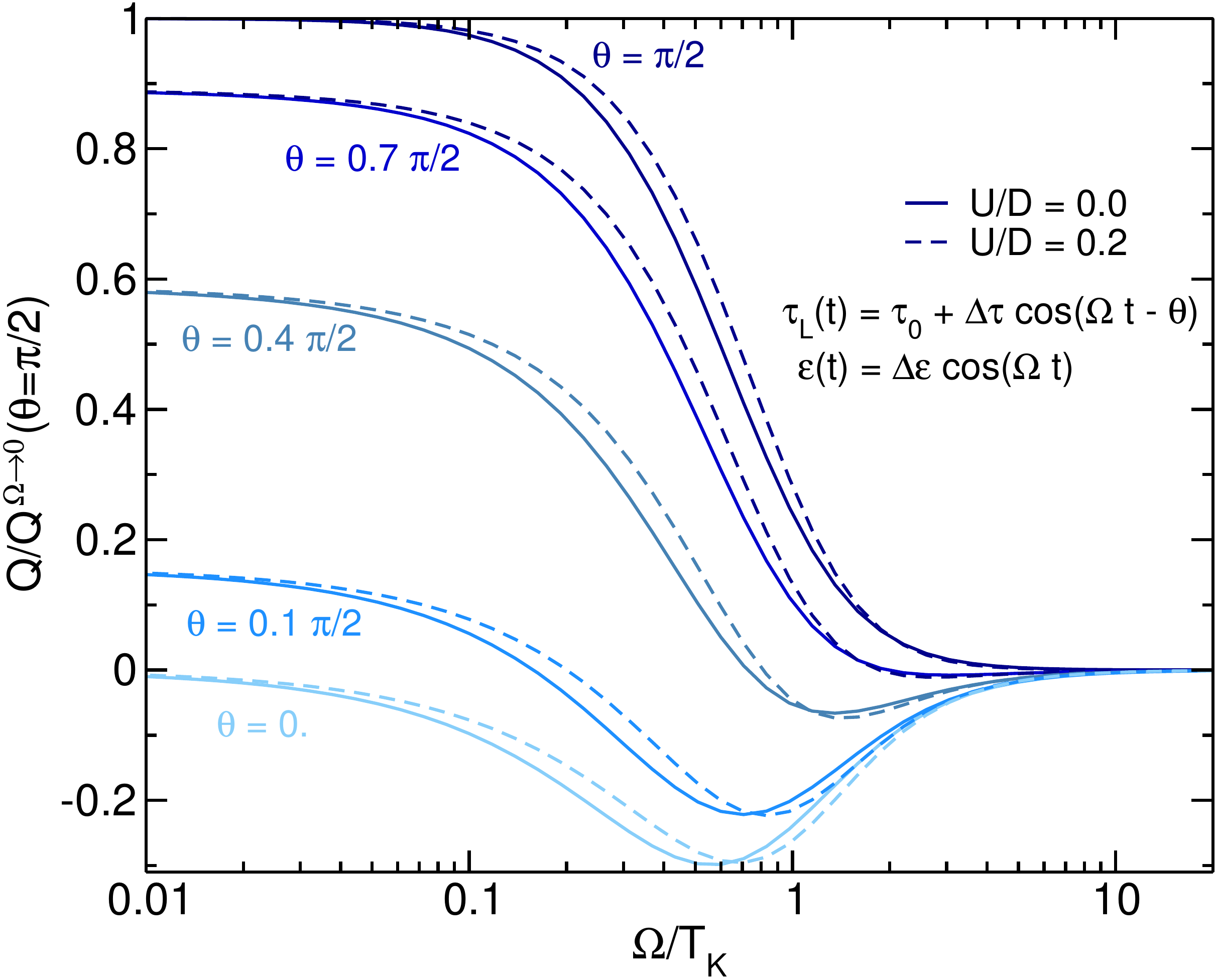}
\caption{Pumped charge for various phase differences $\theta$ between the signals of left hopping and onsite energy with $\Delta \tau /\tau_0 = \Delta \epsilon/T_{\rm K} = 0.05$ for $U/D = 0.0, 0.2$ and accordingly $T_{\rm K}/D = 2.5 \cdot 10^{-5} ,7.93 \cdot 10^{-5}$. A monotonously decreasing function for $\theta = \pi/2$ with increasing driving frequency becomes a non-monotonous function with maximal pumping in the opposite direction as the phase difference vanishes. The behavior is independent of interaction. In the limit of fast driving no charge is pumped at all, independent of the phase difference between the signals.}
\label{fig:qpumphi}
\end{figure}

In the traditional pump setup\cite{Brouwer98} two parameters are varied time periodically in the adiabatic limit. Keeping a finite phase shift between their signals leads to pumped charge from one reservoir into the other. As discussed for protocol 4, we realize this setup by varying the left hopping on the dot and the onsite energy around the particle-hole symmetric point. Maximal charge is pumped from the left to the right reservoir, if the signal of the left hopping is retarded by $\pi/2$ compared to the signal of the onsite energy. We thus define a positive pumped charge $Q$ if it flows from the left to the right reservoir and consequentially negative if charge is pumped in the opposite direction.

We want to examine a gradual evolution from this case into an in-phase quantum pump by decreasing the phase difference between the signals of the two periodically driven parameters. Starting with a maximal phase difference $\theta = \pi/2$, $\theta$ is decreased gradually and the pumped charge $Q$ is depicted for the whole range of possible driving frequency in Fig.\,\ref{fig:qpumphi}.
 For $\theta = 0$  we end up in an in-phase quantum pump with two time periodic parameters, which are varied with the same signal, i.e.\,oscillate in phase.
Independent of the phase difference, no charge is pumped in the limit of large driving frequency,  already at $\Omega = 10 T_{\rm K}$ a negligibly small pumped charge is obtained.
For the maximal phase difference $\theta = \pi/2$, $Q$ is maximal in the adiabatic limit and decreases monotonously and rapidly as the driving frequency $\Omega$ approaches the low energy scale $T_{\rm K}$. As $\theta$ decreases, the function becomes non-monotonic and establishes a minimum with a negative sign, such that for $\theta=0$, maximal charge is pumped in the opposite direction for a moderate driving frequency of $\Omega \lesssim T_{\rm K} $.\cite{Croy2012b} The pumped charge vanishes in either the adiabatic or in the antiadiabtic limit. The effect of the interaction on this behavior is marginal for all phase differences (besides the renormalization of the low energy scale $T_{\rm K}$).

 The in-phase quantum pump is sometimes also called single parameter quantum pump in the literature,\cite{Cavaliere2009} since it might be realized by one periodically varying external field controlled by a single gate voltage experimentally.\cite{Kaestner2015} Here we use a different nomenclature. For us a single parameter pump is defined by having only one time periodic model parameter.
 It will be discussed in the next section.  

\subsection{Single parameter quantum pump}
\label{sec:singleparpump}
In section \ref{sec:prot1} we have discussed the renormalization of the higher harmonics of the left hopping, time periodically varied with an arbitrary signal. We want to discuss now how the observed power law in $k\Omega$ in the limit of large driving is reflected in an observable.  
For this we concentrate on a simple sine signal
\begin{equation}
\tau_L = \tau_0 + \Delta \tau \sin(\Omega t)
\end{equation}
in the small amplitude limit. We break particle hole symmetry, by choosing a finite (but small compared to $T_{\rm K}$) onsite energy $\epsilon_0 = 0.4 T_{\rm K}$. 
The resulting single parameter pump has a finite mean current $J_{k=0}$ without applied bias voltage.

The mean current $J_{k=0}$ is computed in a non-interacting, effective model with renormalized parameters which incorporate all correlation physics. The analytic expression for the dc current $J_0$ is obtained by interpreting the Floquet index as an extra spatial index \cite{Shirley1965, Gomez2013} as already discussed for protocol 3 (replica idea). Identifying the leading order contributions to the current leaving the left reservoir, shows that the system can be restricted to an effective three terminal setup as the finite current is a consequence of temporary excursions of the electron in the $k=\pm 1$ replicas of the system.
 Using a Landauer-B\"uttiker formula, \cite{Buttiker1986, Landauer1996} we obtain an analytic expression of the mean current (for more details see the supplemental material of Ref.\,\onlinecite{Eissing16}), which reads
\begin{equation}
 J_{\text{L},k=0}\stackrel{\Omega\gg\epsilon}{=}\frac{1}{2\pi}\left(\frac{|\tau^{{\rm ren}}_{{\rm L},1|}}{\tau^{\rm ren}_{{\rm L},0} }\right)^2 
T_{\rm K}\arctan\left(\frac{2\epsilon}{T_{\rm K}}\right)
 \end{equation}
 in the limit of large driving frequency $\Omega$ and describes the exact solution in case of vanishing interaction.
Due to the leading dependency on $\tau_{{\rm L},k=1}^2$, the power law in the driving frequency of the higher harmonics then manifests in   
\begin{equation}
\frac{J_{\rm L,k=0}}{T_{\rm K}} \sim \left( \frac{\Omega}{T_{\rm K}}\right)^{-2U/(\pi D)} .
\end{equation}
In Fig.\,\ref{fig:exponenten} the exponent $\alpha_J = U/(\pi D) + \mathcal{O}(U^2)$ is depicted. It is obtained via a logarithmic derivative $d\ln(J_{\text{L},k=0})/d\ln(\Omega)$ from the numerically calculated current via Eq.\,(\ref{eq:cur}).

\begin{figure}
\includegraphics[width=\columnwidth]{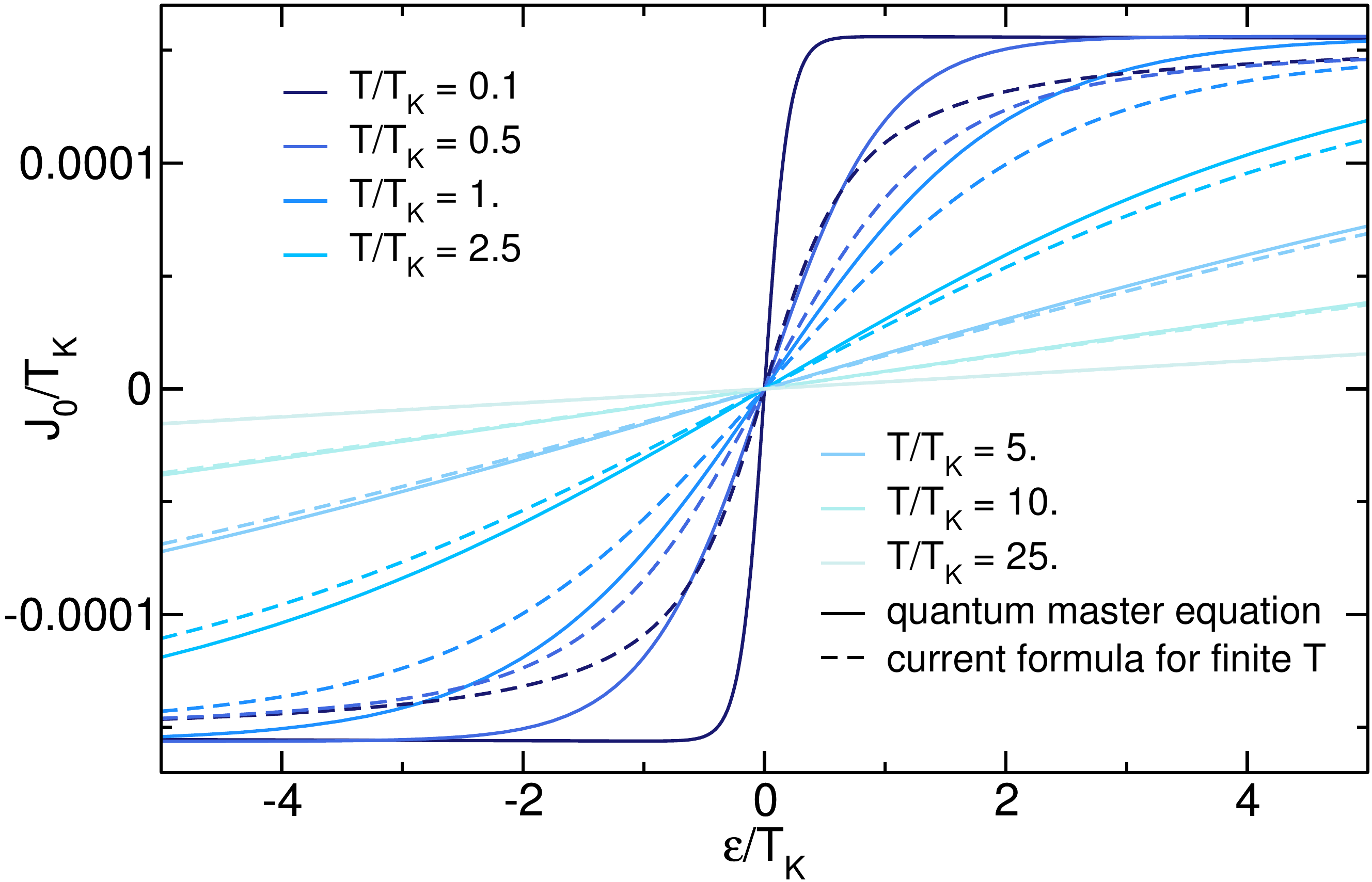}
\caption{Mean left current as a function of onsite energy $\epsilon$ for protocol 1 with the non-interacting $T_K = 0.01$. Only the left hopping element is varied time periodically with a sine signal in the non-interacting system. Results obtained with the quantum master equations are compared to the Landauer- B\"uttiker type formula extended to finite temperature.  The quantum master equation is capable to reproduce the finite mean current for large driving frequency $\Omega/(2\pi T_{\rm K}) = 200$. With increasing temperature the agreement with the analytic formula improves.}
\label{fig:MarkovCurL}
\end{figure}

This finite current cannot be obtained in methods, which rely on tunneling rates with a single time argument as e.g.\,used in Ref.\,\onlinecite{Cavaliere2009} for the regime $\Omega \lesssim \Gamma$. 
A different approach was put forward in Ref.\,\onlinecite{Braun2008}.
The authors take care of the two-time structure of tunneling in and out, but integrate out one time argument using time scale separation in the anti-adiabatic limit.
This approach allows to compute a finite current in a spinful single level setup in the anti-adiabatic limit for the same protocol as considered here.
 
In order to relate to these results and understand our setup in the context of quantum master equations, we take explicitly advantage of the time periodicity. We set up master equations in Floquet space
\begin{align}
\dot{p}_s (t) &=\sum_{ks'} e^{-ik\Omega t} W^k_{ss'} p_s' (t), \nonumber \\
\left< I_\gamma\right> (t) &=  \sum_{knss'} e^{-i(k+n)\Omega t} W^{k,\gamma}_{ss'} p^n_{s'}
\end{align}
for the non-interacting system with only time periodic hopping elements and compute the mean value of the left current
\begin{align}
\left< I_{{\rm L},0}\right>  &=  \sum_{nss'}  W^{-n,L}_{ss'} p^n_{s'} .
\end{align}
The kernel $W^k_{ss'} $ is computed to the first order in tunneling coupling in Floquet space (more details of the calculation are presented in the appendix).

These results (solid lines) are compared in Fig.\,\ref{fig:MarkovCurL} to the analytic formula (dashed line) obtained via Landauer - B\"uttiker formalism in the replica picture (expression can be found in the appendix), which can be extended to finite temperature as long as $T < \Omega$.  The left mean current as a function of onsite energy is displayed.
The quantum master equation is indeed able to reproduce the finite mean current with a correct qualitative behavior. The perturbative expansion in the tunneling coupling improves with increasing temperature leading to a good agreement in the regime $T >T_{\rm K}$.
 
\section{Conclusion}
We developed a FRG approach for the long time behavior of time periodic, interacting quantum dot setups. The method takes advantage of the periodicity explicitly and uses Floquet-Green's functions to set up the functional RG in Floquet space. The approach in the natural basis for time periodic problems thus requires no further restrictions on the driving frequency or amplitude and allows to study the role of the driving frequency as a cutoff scale (also analytically in the limit of small driving amplitude).

 We applied the approach to the interacting resonant level model as a prototype model for quantum dots dominated by charge fluctuations. Four protocols are examined with different combinations of time periodic hopping elements and onsite energy in the limit of small driving amplitudes. The decoupling of the renormalization of $k \neq 0$ channel of the hopping elements and the onsite energy in this limit allows for an analytical description of their renormalization to the leading order in $U, \frac{1}{D}, p=\frac{\tau_{k \neq 0}}{\tau_0} = \frac{\epsilon_{k \neq 0}}{T_{\rm K}}$ at the particle hole symmetric point besides the full numerical solution. It shows that the $k$th component only depends on the driving frequency with the prefactor $k$, i.e. only on $k \Omega$. This renders the mean value independent of the driving frequency and the exact protocol of time periodicity.
More importantly, it reveals a new power law for the higher harmonics of the time periodic hopping $\tau_{k\neq 0}$ in the driving frequency $k\Omega$ with a $U$ dependent exponent in a setup with only time dependent hopping elements. This $k$ and $\Omega$ dependent renormalization of $\tau_{k\neq 0}$ has interesting effects on the renormalized time-periodic signal: its shape is changed and the amplitude is rectified or amplified depending on the sign of the interaction. 
On the other hand the renormalization of $\tau_{k\neq0}$ in a setup of periodically varied onsite energy does not show the usual dependence on a single infrared cutoff, but a complicated interplay of all involved energy scales and thus does not show power-law behavior.
 
These results are complemented by a consideration of a setup with a time periodic onsite energy in the whole range of driving amplitude. Here the effective reservoir distribution function is tuned by choosing a certain ratio of driving amplitude and frequency. Knowledge of the form of the effective reservoir distribution function allows to analyse how the edges of the reservoir distribution functions at $n\Omega$ (which lead to infrared divergencies) are reflected in the renormalization group flow. A reservoir distribution function can be constructed in a time periodic setup which resembles the one of an applied bias voltage and thus leads to similar renormalization of the hopping element.
 
Finally, we made use of the accessibility of the whole range of driving frequency and discussed the pumped charge of in-phase quantum pumps and the finite mean current $J_0$ of a single parameter pump, which directly reflects the power law of the higher harmonics of the time periodic hopping.

The presented Floquet FRG can be extended to a higher order truncation scheme of the flow equations. A more general transformation to Floquet space\cite{Tsuji2008} can be employed to treat the four time dependent two-particle vertex function efficiently, but the numerical effort to include the arising full frequency dependency as well as an appropriate number of higher harmonics renders this a much more involved task.
Besides of the study conducted here, describing periodically driven one-dimensional lattices provides another promising application of the developed formalism. The FRG has been vital in understanding, e.g., the boundary and impurity physics of Luttinger liquids in \cite{Meden2008, Metzner2012} and out-of \cite{Jakobs2007} equilibrium. This will be the subject of a forthcoming publication.

\section*{Acknowledgement}
We thank T.\,Pl\"ucker, J.\,Splettst\"o\ss er, T.\,Suzuki and H.\,Schoeller for discussions. This work was supported by the Deutsche Forschungsgemeinschaft (RTG 1995 and DFG KE 2115/1-1).

\newpage
\onecolumngrid
\vspace{\columnsep}
\section*{Appendix: Details of the Analytic calculations}
\label{sec:appendix}

\noindent Here details of the analytic calculations are presented. The main goal is to derive an analytic expression for the renormalized parameters in the time periodic interacting resonant level model and this way gain an understanding of the underlying renormalization flow of our FRG approach in the limit of small amplitudes $\Delta \tau, \Delta \epsilon$. We define a dimensionless parameter $p = \frac{\tau_{k \neq 0}}{\tau_0} = \frac{\epsilon_{k \neq 0}}{T_{\rm K}}$, which is kept small $p \ll 1$ for all calculations.

\phantom{x}
\noindent The inverse of the retarded reservoir dressed dot Green's function in Floquet space is defined as
\begin{equation}
(\dunderline{G}^{\text{ret},\Lambda})^{-1} = 
\begin{pmatrix}
\hat{H}_{-1 -1} & \hat{H}_{-10} & 0 \\
\hat{H}_{0-1} & \hat{H}_{00} &\hat{H}_{01}   \\
 0 & \hat{H}_{10}  & \hat{H}_{11}  \\
 \end{pmatrix},
 \label{Gretinv}
\end{equation}
where only Floquet indices $k,k'$ are shown, while each $\hat{H}_{k,k'}$ is itself a matrix in real space. The infinite Fourier space is already truncated after the first higher harmonic i.e.\,only the subspace spanned by $k=0, \pm 1$ is included. This is a consistent approximation to $\mathcal{O}(p)$ if we focus on the renormalization of the $k=0,\pm1$ coefficients. 

  \phantom{x}
\noindent The diagonal elements in the Fourier space $\hat{H}_{k,k}$ are the $k=0$ components of the 'reservoir dressed' Floquet Hamiltonian defined in Eq.\,(\ref{eq:FloqH})
\begin{equation}
\hat{H}_{\rm k,k} = 
\begin{pmatrix}
  \omega + k\Omega + i(D+ \Lambda) & \tau_{\rm L,0} & 0   \\
  \tau^*_{\rm L,0} & \omega + k\Omega - \epsilon_0+ i \Lambda & \tau_{\rm R,0} \\
 0  & \tau^*_{\rm R,0} &  \omega + k\Omega + i(D+ \Lambda) \\
 \end{pmatrix},
\end{equation}
the respective offdiagonal elements $\hat{H}_{k,k'}$ for $k\neq k'$ are 
\begin{equation}
\hat{H}_{\rm k,k'} = 
\begin{pmatrix}
0& \tau_{{\rm L},k'-k} & 0   \\
 \tau_{{\rm L},k'-k} & -\epsilon_{k'-k} &  \tau_{{\rm R},k'-k} \\
 0  &  \tau_{{\rm R},k'-k}&  0\\
 \end{pmatrix},
 \end{equation}
 with the corresponding $k'-k$ Fourier coefficients as defined in Eq.\,(\ref{eq:e_eff_coeff}). As discussed in the main text, we set $\tau_{{\rm L},k=0} = \tau_{{\rm R},k=0} = \tau_0$ and $\epsilon_{k=0} = 0$.
 
\phantom{x}
\noindent We are only interested in the leading order of the small parameter $p$ and thus neglect all terms $\mathcal{O}(p^2)$. Then the retarded Green's function is given by
 \begin{equation}
\dunderline{G}^{\text{ret},\Lambda} = 
\begin{pmatrix}
 \hat{H}^{-1}_{-1-1} & - \hat{H}^{-1}_{-1-1} \hat{H}_{01} \hat{H}^{-1}_{00}& 0 \\
- \hat{H}^{-1}_{00}\hat{H}_{10}\hat{H}^{-1}_{-1-1} & \hat{H}^{-1}_{00} & - \hat{H}^{-1}_{00} \hat{H}_{01}   \hat{H}^{-1}_{11} \\
 0& - \hat{H}^{-1}_{11} \hat{H}_{10} \hat{H}^{-1}_{00}  & \hat{H}^{-1}_{11}  \\
 \end{pmatrix},
 \label{eq:Gretstructure}
\end{equation}  
with only the Floquet indices shown and summation over the quantum numbers of the real space assumed.
The inverse of the Hamiltonian in real space is then given by
\begin{align}
\mbox{\small $\hat{H}^{-1}_{\rm k,k} = \frac{1}{K} \begin{pmatrix}
  (\omega + k\Omega + i(\Lambda + D)) (\omega + k \Omega + i \Lambda) - |\tau_{0}|^2 & \tau_{0} (\omega +k \Omega +i(D+ \Lambda)) & \tau_{0} \tau_{0}   \\
  \tau^*_{0} (\omega +k \Omega +i(\Lambda+D)) & (\omega + k\Omega + i (\Lambda + D))^2 & \tau_{0} (\omega+ k\Omega + i(D+ \Lambda)) \\
 \tau_{0}^* \tau_{0}^*  & \tau_{0}^*( \omega + k\Omega + i(D+ \Lambda)) & (\omega+ k\Omega + i(D + \Lambda))(\omega + k\Omega + i\Lambda)- |\tau_{0}|^2 \\
 \end{pmatrix}$ } ,
\end{align} 
with
\begin{equation}
K = (\omega+ k\Omega + i(D + \Lambda))^2(\omega + k\Omega + i \Lambda)-2|\tau_{0}|^2 (\omega + k\Omega+ i(D + \Lambda)).
\end{equation}
Note the transparent structure of $\dunderline{G}^{\rm ret, \Lambda}$ in the Fourier space in Eq.\,(\ref{eq:Gretstructure}) to the leading order of $\mathcal{O}(p)$: The diagonal elements only feature the inverse of the respective Hamiltonian entry. The off-diagonal elements with an effective (or physical) Fourier coefficient $k_1 = k-k' \neq 0$, depend on the $k_1$th coefficient of the Hamiltonian and $k$ diagonal elements of the inverse Hamiltonian. 

\phantom{x}
\noindent The Keldysh reservoir self-energy is given in the $k=0, \pm 1$ Fourier space as
\begin{equation}
\dunderline{\Sigma}^{\text{K}} = 
\begin{pmatrix}
\hat{\Sigma}^{K}_{-1-1} & 0 &0 \\
0 &  \hat{\Sigma}^{K}_{00} & 0  \\
 0  & 0 & \hat{\Sigma}^{K}_{11}  \\
 \end{pmatrix},
\end{equation}
where each $\hat{\Sigma}^{\rm K}_{kk'}$ is a matrix in $\mathcal{R}$
\begin{equation}
\hat{\Sigma}_{kk}^{\text{K}} = 4 i D
\begin{pmatrix}
  \theta [-(\omega + k \Omega)] -\frac{1}{2}& 0  &0\\
  0 & 0 & 0 \\
 0 &0  & \theta [-(\omega + k \Omega)]-\frac{1}{2} \\
 \end{pmatrix}.
\end{equation}
Using the Dyson equation to compute the Keldysh Green's function (for reservoirs that fulfill the dissipation-fluctuation-theorem) \cite{DissCK}
\begin{equation}
\dunderline{G}^{K, \Lambda} = \dunderline{G}^{\rm ret, \Lambda} \dunderline{\Sigma} ^{K} \dunderline{G}^{\rm adv, \Lambda}
\end{equation}
the Keldysh Green's function becomes to linear order in $p$
\begin{equation}
\dunderline{G}^{\text{K},\Lambda} = 
\begin{pmatrix}
\hat{G}^K_{-1-1} & \hat{G}^K_{-10} & 0 \\
\hat{G}^K_{0-1} & \hat{G}^K_{00} & \hat{G}^K_{01} \\
0 & \hat{G}^K_{10} & \hat{G}^K_{11} \\
\end{pmatrix}
 \label{eq:GKL}
\end{equation}
with
\begin{align*}
 &\hat{G}^{\rm K}_{-1-1} = \hat{H}^{-1}_{-1-1} \hat{\Sigma}^K_{-1-1} (\hat{H}^{-1}_{-1-1})^*  \\
 &\hat{G}^{\rm K}_{-10} = -\hat{H}^{-1}_{-1-1}\hat{\Sigma}^K_{-1-1} (\hat{H}_{00}^{-1})^* \hat{H}_{01}  (\hat{H}^{-1}_{-1-1})^* -\hat{H}^{-1}_{-1-1} \hat{H}_{01} \hat{H}_{00}^{-1} \hat{\Sigma}^K_{00}  (\hat{H}^{-1}_{00})^*\\  
 &\hat{G}^{\rm K}_{0-1} = -\hat{H}^{-1}_{00}\hat{\Sigma}^K_{00} (\hat{H}_{-1-1}^{-1})^* \hat{H}_{10}  (\hat{H}^{-1}_{00})^* -\hat{H}^{-1}_{00} \hat{H}_{10} \hat{H}_{-1-1}^{-1} \hat{\Sigma}^K_{-1-1}  (\hat{H}^{-1}_{-1-1})^*\\
 &\hat{G}^{\rm K}_{00} = \hat{H}^{-1}_{00} \hat{\Sigma}^K_{00} (\hat{H}^{-1}_{00})^* \\
 &\hat{G}^{\rm K}_{01} = -\hat{H}^{-1}_{00}\hat{\Sigma}^K_{00} (\hat{H}_{11}^{-1})^* \hat{H}_{01}  (\hat{H}^{-1}_{00})^* -\hat{H}^{-1}_{00} \hat{H}_{01} \hat{H}_{11}^{-1} \hat{\Sigma}^K_{11}  (\hat{H}^{-1}_{11})^*\\
&\hat{G}^{\rm K}_{10} =-\hat{H}^{-1}_{11}\hat{\Sigma}^K_{11} (\hat{H}_{00}^{-1})^* \hat{H}_{10}  (\hat{H}^{-1}_{11})^* -\hat{H}^{-1}_{11} \hat{H}_{10} \hat{H}_{00}^{-1} \hat{\Sigma}^K_{00}  (\hat{H}^{-1}_{00})^* \\
&\hat{G}^{\rm K}_{11} =  \hat{H}_{11} \hat{\Sigma}^K_{11} (\hat{H}^{-1}_{11} )^* 
\end{align*}
and $\hat{H}^{-1}_{k,k}$ defined as above. Again summation over suppressed real space quantum numbers is assumed.  
The structure of the elements is transparent and allows to generalize the expression to
\begin{equation}
G^{\rm K}_{kk} = \hat{H}^{-1}_{kk} \hat{\Sigma}^K_{kk} (\hat{H}^{-1}_{kk})^* ,
\end{equation}
\begin{equation}
\hat{G}^{\rm K}_{kk'} =-\hat{H}^{-1}_{kk}\hat{\Sigma}^K_{kk} \hat{H}_{k'k'}^{-1,*} \hat{H}_{kk'}  \hat{H}^{-1,*}_{kk} -\hat{H}^{-1}_{kk} \hat{H}_{kk'} \hat{H}_{k'k'}^{-1} \hat{\Sigma}^K_{k'k'}  \hat{H}^{-1,*}_{k'k'} 
\end{equation}
for all $k\neq k'$. Let us emphasize that the diagonal elements of $G^{\rm K}$ only depend on diagonal elements with the same $kk$ indices, while the off-diagonal elements depend on diagonal elements with coefficients $kk$ or $k'k'$ and only on off-diagonal elements with the same Fourier coefficients. Since any other contributions of a different higher harmonic would be of higher order in $p$, the different $k$ channels decouple and hence can be considered independently.   As a result, these expressions can be generalized from $k=0,\pm1$ to the whole Fourier space. 

\phantom{x}
\noindent These approximate $\hat{G}^{\rm K}_{kk'}$ are inserted in the right hand side of the full flow equations (Eq.\,(\ref{eq:flowtau}) and (\ref{eq:floweps}) of the main text)
\begin{align*}
\partial_{\Lambda} \tau^{\Lambda}_{\rm L(R),k} &= -\frac{U}{4\pi i}\, \partial^{*}_{\Lambda} \int d\omega\, G^{\rm K,\Lambda}_{12(23);0k} (\omega) , \\
\partial_{\Lambda} \epsilon^{\Lambda}_{k} &= -\frac{Ui}{4\pi}\, \partial^{*}_{\Lambda} \int d\omega\, \left( G^{\rm K,\Lambda}_{11;0k} (\omega) + G^{\rm K,\Lambda}_{33;0k} (\omega) \right)
\end{align*}
 to compute the renormalization analytically in the leading order of $\mathcal{O}(p)$.

\subsection{Renormalization of the $k=0$ channel}
\noindent  For the renormalization of the $k=0$ components of the hopping elements we can consider any of the diagonal entries of $G^{\rm K}$,
which are independent of the driving frequency and of any higher harmonic.
As a consequence the renormalization of the $k=0$ component is independent of the exact driving protocol.
 
We thus concentrate on the mean values of the hopping matrix element only, since the mean value of the onsite energy is not renormalized at the particle hole symmetric point (see main text).
We include all contributions to the order of $\frac{1}{D}$ of the matrix entries
\begin{align}
G^{\rm K}_{12(23),00} = \hat{H}^{-1}_{1(2)l,00} \hat{\Sigma}^K_{ll,00} \hat{H}^{-1,*}_{l2(3),00},
\end{align}
which results in
\begin{equation}
\partial_{\Lambda} \tau_0 = \partial^*_{\Lambda} \int d\omega \frac{Ui}{4\pi} \frac{\tau^{\Lambda}_0 (\omega+i\Lambda)}{\omega+i(\Lambda+D)(\omega+i\Lambda)-2|\tau_0|^2} \frac{1}{(\omega-i(\Lambda + D))(\omega-i\Lambda)-2 |\tau_0|^2}  4iD\left[\theta (-\omega)- \frac{1}{2}\right].
\end{equation}
In order to rewrite the star derivative, which denotes a derivative of only one bare Green's function, $\tau_0$ in the denominator is set to its initial value and $\tau_0^{\Lambda}$ is moved in front of the derivative, which allows to substitute $\partial^*_{\Lambda}$ by  $\partial_{\Lambda}$
\begin{align}
\partial_{\Lambda} \tau_0 =  - \frac{U}{\pi D} \frac{\tau^{\Lambda}_0}{D} \partial_{\Lambda} \int d\omega \frac{(\omega+i\Lambda)/D}{(\omega+i(\Lambda+D))/D(\omega+i\Lambda)/D-2|\tau_0|^2/D^2}  \frac{ \left[\theta (-\omega)- 1/2 \right]}{(\omega-i(\Lambda + D))/D(\omega-i\Lambda)/D-2 |\tau_0|^2/D^2} ,
\end{align}
where all parameters are divided by $D$. Including all contributions to the order of $\frac{1}{D^2}$ in the denominator the expression can be computed to
 \begin{equation}
\partial_{\Lambda} \tau^{\Lambda}_{0}= - \frac{U}{\pi D} \frac{\tau^{\Lambda}_{0}/D}{(\Lambda/D)^2 + \Lambda/D + 2(\tau_{0}/D)^2}
 \end{equation}
 with $\tau^{\Lambda = \infty}_0 = \tau_0$ as initial value, reproducing the differential equation for the time independent equilibrium setup. \cite{Karrasch2010}
Solving the differential equation analytically, results in
\begin{align}
\frac{\tau_{0}^{\rm ren}}{\tau_{0}} &\stackrel{\hspace{8mm}}{=}
\left[ \frac{1-\sqrt{1-8(\tau_0/D)^2}}{1+\sqrt{1-8(\tau_0/D)^2}}\right]
^{-\frac{U}{\pi D}[1-8(\tau_0/D)^2]^{-1/2}}\notag\\
&\stackrel{D \gg \tau_0}{=} \left(\frac{2\tau_0^2}{D^2}\right)^{-\frac{U}{\pi D}} \label{eq:ren1}
\end{align}
i.e.\,the same power law as it has been discussed in Sect.\,\ref{sec:nonequIRLM} for the time independent IRLM. 

In the following, we compute the analytic expressions for the higher harmonics in the following initially with the unrenormalized value $\tau_0$ and subsequently use the Eqs.\,(\ref{eq:threnk0}) to feedback the renormalized values. 

\subsection{Protocol 1: Renormalization of higher harmonics: $k\neq 0$}

\noindent In protocol 1 the left hopping is time periodically varied, setting $\tau_k \neq 0$, while  $\epsilon_ {k} = \tau_{{\rm R},k} = 0$.
 To describe the renormalization of $\tau_{{\rm L},k}$, the flow equation
\begin{align}
\partial_{\Lambda} \tau^{\Lambda}_{{\rm L},k} =&  \frac{Ui}{4\pi}\, \partial^{*}_{\Lambda} \int d\omega \hspace{2mm} G^{\rm K,\Lambda}_{12;0k} (\omega)
\end{align}
is considered. 
Summing up all contributions of the leading order in $\frac{1}{D}$, results in
 \begin{align}
 G_{12,0k}^{\rm K, \Lambda} (\omega) =\hspace{5mm} &\frac{\tau^{\Lambda}_{\rm L,k} (\omega +k\Omega - i(\Lambda+D))}{(\omega+ k\Omega -i(D+ \Lambda)) (\omega+k\Omega - i\Lambda) - 2|\tau_{0}|^2} \nonumber \\ &\frac{(\omega - i\Lambda)}{(\omega - i(D+ \Lambda))(\omega-i\Lambda) - 2|\tau_{0}|^2}   \frac{4iD \left[\theta (-\omega)- 1/2\right](\omega  + i\Lambda)}{(\omega + i(D + \Lambda)) (\omega+i\Lambda) - 2|\tau_{0}|^2} 
\end{align}
again including contributions to the order of $1/D^2$ and rewriting the star derivative leads to
\begin{align}
\partial_{\Lambda} \tau^{\Lambda}_{{\rm L},k \neq 0} = - \frac{U}{\pi D} \frac{\tau^{\Lambda}_{L,k}}{D} \partial_{\Lambda} \int d\omega \frac{(\omega - i\Lambda)/D \,[\theta (-\omega) - 1/2]}{-\frac{\omega + k\Omega - i\Lambda}{D} \frac{\omega - i\Lambda}{D} + \frac{2|\tau_{0}|^2}{D^2} i (\frac{\omega + k\Omega- i\Lambda}{D} + \frac{\omega - i\Lambda}{D}) +\frac{2|\tau_{0}|^4}{D^4} } .
\end{align}
The integral on the right hand side can be performed
\begin{align}
\partial_{\Lambda} \tau^{\Lambda}_{L,k \neq 0} =  - \frac{U}{\pi D}  \frac{ \tau^{\Lambda}_{\text{L},k} \Lambda/D^2}{\frac{\Lambda^2}{D^2}+ (4 \frac{|\tau_0|^2}{D^2}+ \frac{ik\Omega}{D}) \frac{\Lambda}{D}+ \frac{2i|\tau_0|^2 k \Omega}{D^3}+\frac{4|\tau_0|^4}{D^4}} .
\end{align}
This describes the flow of the higher harmonics in protocol 1.
These differential equations can be solved analytically, yielding
\begin{align}
\frac{\tau_{\text{L},k \neq 0}^{\rm ren}}{\tau_{\text{L},k\neq
0}}
\stackrel{\hspace{8mm}}{=}& e^{\frac{U}{2\pi D
k\Omega}(2ik\Omega + \frac{4
|\tau_0|^2}{D}) \arctan\left(\frac{k\Omega}{D+ 2 |\tau_0|^2/D}\right)}
 \left(D+ \frac{2|\tau_0|^2}{D}\right)^{ \frac{2iU|\tau_0|^2}{\pi D^2(k\Omega)}}
\left[k^2\Omega^2 + \left(D +
\frac{2|\tau_0|^2}{D}\right)^2\right]^{\frac{U(k\Omega
-2i|\tau_0|^2/D)}{2\pi D k\Omega}} \notag\\
&\times e^{\frac{-U}{2\pi D k\Omega
}(2ik\Omega + 4 \frac{|\tau_0|^2}{D})
\arctan\left(\frac{k\Omega}{2|\tau_0|^2/D}\right)}
\left(\frac{2|\tau_0|^2}{D}\right)^{\frac{-2iU|\tau_0|^2}{\pi D^2 k\Omega}}
\left[k^2 \Omega^2 +\left(\frac{2|\tau_0|^2}{D}\right)^2\right]^{
\frac{-U(k\Omega
-2i|\tau_0|^2/D)}{2\pi D k\Omega} } \notag\\
\stackrel{D \gg \tau_k}{=}&   e^{\frac{-U}{2 \pi D k\Omega} \frac{4
|\tau_0|^2}{D} \arctan\left(\frac{
k\Omega}{2 |\tau_0|^2/D}\right)} \left[ \frac{k^2\Omega^2+
4(|\tau_0|^2/D)^2}{D^2}\right] ^{-\frac{U}{2\pi D}}
e^{-iU/(\pi D)\arctan\left(\frac{k\Omega}{2|\tau_0|^2/D}\right)} \notag \\
&\times \left[\frac{4(|\tau_0|^2/D)^2}{k^2\Omega^2 +4
(|\tau_0|^2/D)^2}\right]^{\frac{U|\tau_0|^2/D}{ \pi D k\Omega}} \notag\\
\stackrel{\Omega \gg T_{\rm K}}{\rightarrow}& \left(
\frac{k\Omega}{D}\right)^{-\frac{U}{\pi D}} \left(i\right)^{-\text{sign}(k)U/(\pi D)}.\label{eq:ren2}
\end{align}
In the last step, we have specified that the frequency $\Omega \gg T_{\rm K}$. 
Substituting $\tau_0$ by its renormalized value, the analytic expression captures the full numerical solution of the flow.

\subsection{Protocol 3: Renormalization of $\epsilon_{k=1}$}
\label{sec:appp3}
\noindent In protocol 3 only the onsite energy $\epsilon$ is time periodically varied. 
We focus on a sinusoidal signal here and choose  $\epsilon(t)=\Delta \epsilon \cos (\Omega t)$.

The flow equation of the first higher harmonic of the onsite energy is 
\begin{align}
\partial_{\Lambda} \epsilon^{\Lambda}_{k=1} &= -\frac{Ui}{4\pi}\, \partial^{*}_{\Lambda} \int d\omega\, \left( G^{\rm K,\Lambda}_{11;01} (\omega) + G^{\rm K,\Lambda}_{33;01} (\omega) \right).
\end{align}
We thus consider 
\begin{align}
&\hat{G}^{\rm K}_{ij,01} = -\hat{H}^{-1}_{il,00}\hat{\Sigma}^K_{ll,00} (\hat{H}_{lm11}^{-1})^* \hat{H}_{mn,01}  (\hat{H}^{-1}_{nj,00})^* -\hat{H}^{-1}_{io,00} \hat{H}_{op,01} \hat{H}_{pq,11}^{-1} \hat{\Sigma}^K_{qq,11}  (\hat{H}^{-1}_{qj,11})^*
\end{align}
at $i=j=1$ and $i=j=3$. All contribution of leading order $\frac{1}{D}$ result in the following flow equation for $\epsilon_{k=1}$
\begin{align}
\partial_{\Lambda} \epsilon_{k=1}^{\Lambda} =\partial_{\Lambda}^* \int d\omega\, \frac{2U}{\pi D}  \frac{\epsilon^{\Lambda}_{k=1}}{D}\left[\frac{(\omega+i\Lambda)/D}{i(\omega+i\Lambda)/D -2|\tau_0|^2/D^2} \frac{|\tau_0|^2/D^2}{-i(\omega+\Omega-i\Lambda)/D- 2|\tau_0|^2/D^2} \frac{\left[\theta(-\omega)-1/2\right]}{-i(\omega-i\Lambda)/D-2|\tau_0|^2/D^2}  \right.& \nonumber \\  + \left.  \frac{|\tau_0|^2/D^2}{i(\omega+i\Lambda)/D- 2|\tau_0|^2/D^2} \frac{ \left[\theta(-\omega-\Omega)-1/2\right] }{i(\omega+\Omega+i\Lambda)/D-2|\tau_0|^2/D^2}\frac{(\omega+\Omega-i\Lambda)/D}{-i(\omega+\Omega-i\Lambda)/D -2|\tau_0|^2/D^2}\right]& . \nonumber \\
\end{align} 

Applying the approximation as discussed for protocol 1 is not possible here. The differential equation cannot be solved as straightforwardly.
If we set all parameters on the right hand side of the flow equation to their respective initial values, the resulting expression can be integrated straight forwardly but only leads to the result of first order perturbation theory.
The resulting expression 
\begin{equation}
\epsilon^{\rm ren}_{k=1} (\Omega) \stackrel{D \gg \tau_k/\Omega}{=} \frac{U}{\pi D} \frac{T_{\rm K}}{2} \frac{\epsilon^{\rm init}_{k=1}}{D}  \frac{i(T_{\rm K}/2 +i \Omega)/D}{\Omega/D (T_{\rm K}+i \Omega)/D} \left[  \ln \left(\frac{T_{\rm K}^2}{T_{\rm K}^2+4\Omega^2} \right) - 2i \arctan\left( \frac{2\Omega}{T_{\rm K}}\right) \right]
\end{equation}
does not encounter any infrared divergencies and is thus sufficient for our purposes. Including the feedback of the $k=0$ channel by substituting $\frac{4\tau^2_0}{D} \rightarrow T_{\rm K}$, the analytic expression captures the numerical solution of the full flow equation.

\subsection{Protocol 3: Renormalization of $\tau_{k=1}$}

\noindent Finally, the renormalization of the left and right hopping is considered, which are equal due to the left right symmetry of the problem. Evaluating $G^{\rm K}_{12(23),01}$ shows that it does not depend on $\tau_{{\rm L(R)},k=1}$ and hence its feedback into its own flow equation is of order $\mathcal{O}(U^2)$ and which is beyond our considerations. 
As a result the renormalization of  $\tau_{{\rm L(R)},k=1}$ is computed in a first order perturbation theory calculation in $U$
\begin{align}
\tau_{{\rm L(R)},k=1} = -U n_{12(23),k=1}  .
\end{align}

\begin{figure}[h]
\begin{minipage}[t!]{0.42\textwidth}
\includegraphics[width=\textwidth]{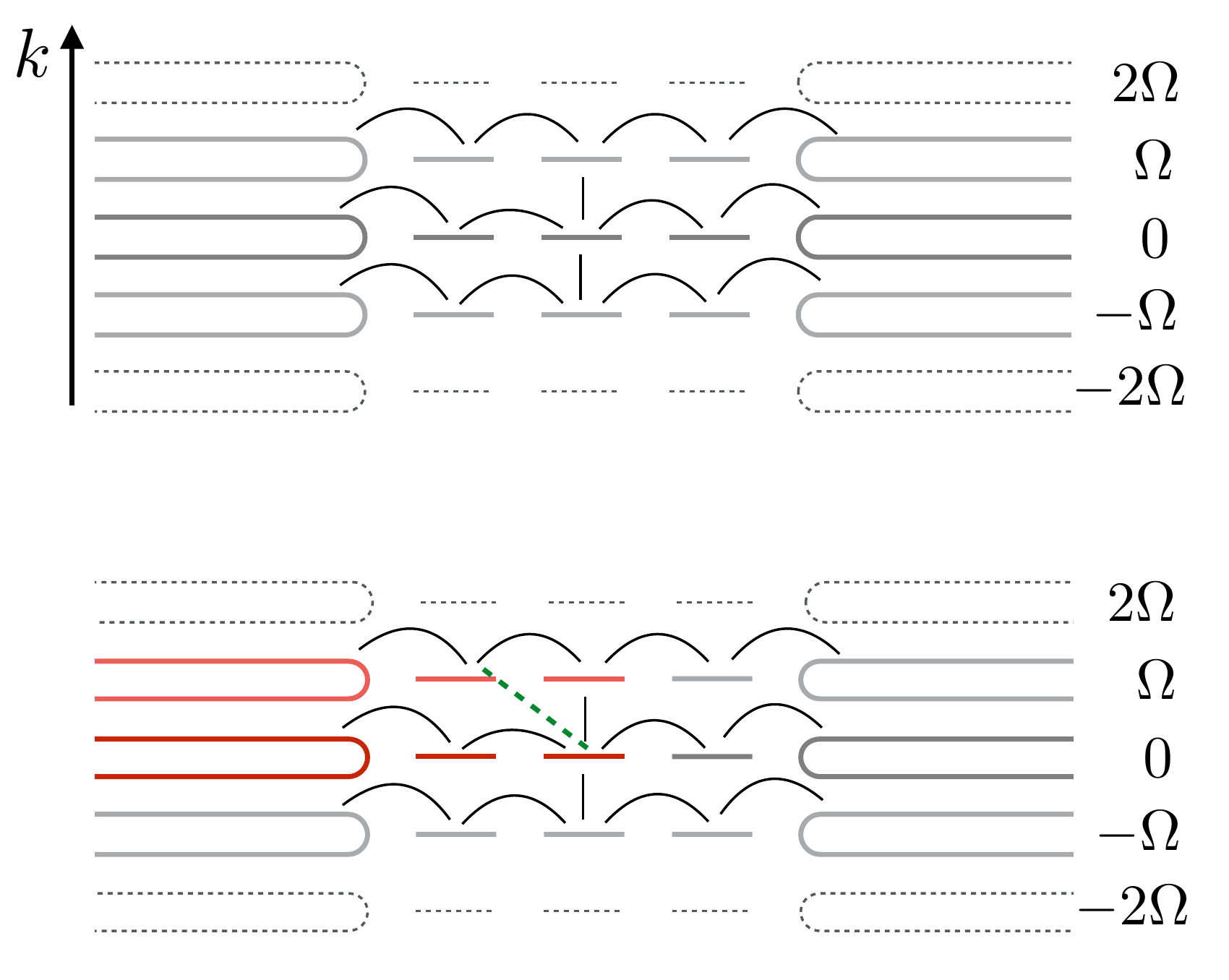}
\end{minipage}
\begin{minipage}[t!]{0.55\textwidth}
\includegraphics[width=\textwidth]{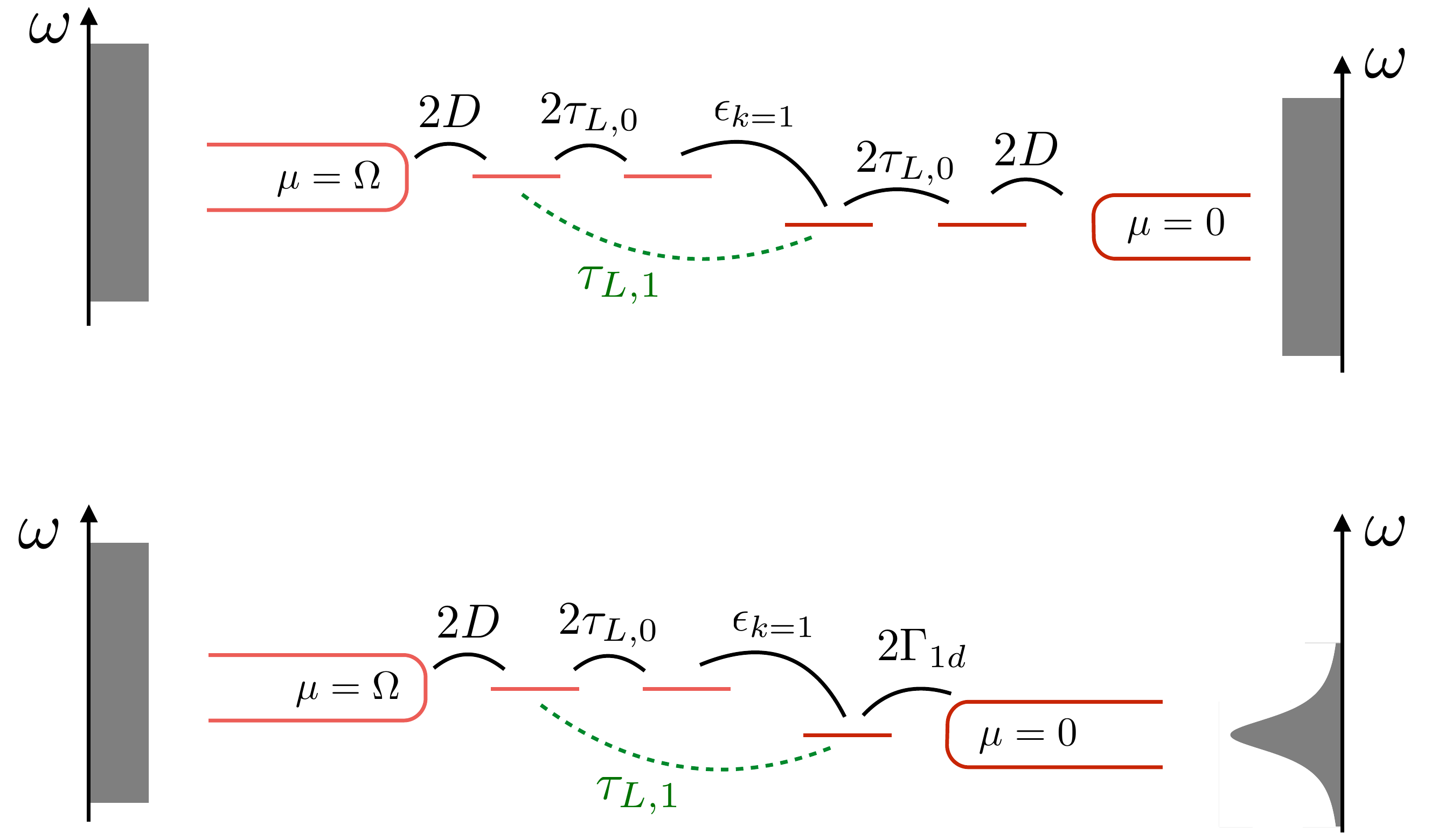}
\end{minipage}
\caption{The time periodically driven 1D system can be understood as a 2D system with replicas shifted by $\Omega$. Neighboring replicas are coupled via $\epsilon_{k=1}$ for the setup with sinusoidally driven onsite energy.  For the calculation of the renormalization of $\tau_{k=1}$ only the zeroth and first replica are of relevance to the order $\mathcal{O}(p)$. The green, dashed line indicates the coupling of interest, which is initially zero.
To realize the compact effective model (as depicted on the right hand side), we take advantage of the left-right symmetry and include the first side into an effective reservoir in the $k=0$ channel.
 In this effective model $\tau^{\rm ren}_{\rm L, k=1}$ is the hopping between the first and third site of the central region as indicated.}
\label{fig:replicaepsonly}
\end{figure}

\phantom{x}
\noindent This contribution is then computed in an effective model, where the Floquet index takes the role of an extra spatial index as discussed in Refs.\,\onlinecite{Shirley1965}, \onlinecite{Zeldovich1967} and \onlinecite{Gomez2013}.
The time periodic system is mapped to a time independent system with an enlarged dimensionality by employing an infinite number of replicas of the system in the auxiliary direction of the Floquet index. The $k$th replica has an effective chemical potential of $\mu = k\Omega$. The various replicas are coupled via the higher harmonics of the time periodic parameters, where the index $k$ indicates the range of the coupling in the auxiliary direction.
 In case of sinusoidal driving of the onsite energy, neighbouring channels are coupled via $\epsilon^{\Lambda}_{k=\pm 1}$. This is depicted in the left, upper panel of Fig.\,\ref{fig:replicaepsonly}.
 
\phantom{x}
\noindent If we focus on the first higher harmonic of the left hopping, we can identify it as the coupling between the first site of the $k=1$ replica and the second site of the $k=0$ channel, which is indicated by a dashed green line. It is zero in the non-interacting system. 
Contributions of the other channels would be of higher order in $p$, we thus focus on the channels $k=0,1$ (marked red in the lower, left panel of Fig.\,\ref{fig:replicaepsonly}). The afore discussed left-right symmetry allows to fold the system with respect to the central site leaving us with a four site model of doubled parameters $2\tau_{\rm L,0}$ and $2D$, respectively. The resulting model is depicted in the right upper panel of Fig.\,\ref{fig:replicaepsonly}. Finally, we incorporate the fourth site of the central region into the right reservoir.  As a result, we have a three site effective model, which is coupled on the one side to a flat band reservoir with the coupling $2D$ and on the other side the central region is coupled with the effective hybridization $2\Gamma_{1d}$ to the right reservoir with an a Lorentzian shaped reservoir distribution function. This is depicted in the lower panel of Fig.\,\ref{fig:replicaepsonly}.

\phantom{x}
\noindent In this effective model, 
\begin{equation}
\tau^{\rm ren}_{\text{L}, k=1} = \tau^{\rm em}_{13} = - U n^{\rm em}_{13}
\end{equation}
with $\tau^{\rm em}/n^{\rm em}$ as hopping/occupation in the effective model.
Thus, only $G^{<}_{13} (\omega)$ needs to be set up in the effective non-interacting model to compute
\begin{equation}
\tau^{\rm ren}_{{\rm L},k=1} = -\frac{U}{2\pi i} \int d\omega\, G^{<}_{13} (\omega)  .
\end{equation}
This becomes
\begin{align}
\int d\omega\,G^{<}_{13} (\omega)  &\stackrel{\hspace{8mm}}{=} \int d\omega\phantom{-} \frac{[(\omega-\Omega)(\omega+2i\Gamma_{1d})-\epsilon_{k=1}^2]2\tau_{L,0}\epsilon_{k=1}[4iD\Theta(-(\omega-\Omega))]}{(\omega- \Omega+2iD) (\omega+2i\Gamma_{1d})(\omega-\Omega)-(2\tau_{L,0})^2(\omega+2i\Gamma_{1d})-\epsilon^2_{k=1}(\omega-\Omega+2iD)} \nonumber \\ &\hspace{22mm}\frac{[(\omega-\Omega-2iD)(\omega-\Omega)-4\tau_{L,0}^2]2\tau_{L,0}\epsilon_{k=1}[4i\Gamma_{1d} \Theta(-\omega)]}{(\omega- \Omega-2iD) (\omega-2i\Gamma_{1d})(\omega-\Omega)-(2\tau_{L,0})^2(\omega-2i\Gamma_{1d})-\epsilon^2_{k=1}(\omega-\Omega-2iD)} \nonumber 
\end{align}
\begin{align}
\stackrel{D \gg \tau_k}{=} \int d\omega\, \Theta(-\omega+\Omega)\frac{4i}{D} \frac{2\tau_{L,0}\epsilon_{k=1}}{D^2} \left[ \frac{\omega-\Omega}{D} \frac{1}{2i(\omega-\Omega)/D-\frac{4\tau_{L,0}^2}{D^2}}\frac{1}{-2i(\omega-\Omega)/D-\frac{4\tau_{L,0}}{D^2}} \right]& \nonumber \\
 \hspace{18mm}+\,\Theta(-\omega) \frac{4i}{D} \frac{\Gamma_{1d}}{D} \frac{2\tau_{L,0}\epsilon_{k=1}}{D^2} \left[  \frac{1}{(\omega-2i\Gamma_{1d})/D} \frac{1}{(\omega+2i\Gamma_{1d})/D} \frac{1}{2i(\omega-\Omega)/D-\frac{4\tau_{L,0}^2}{D^2}} \right]& 
\end{align}
 and integrating results in ($D \gg \tau_k, \Omega$)
 \begin{align}
\tau^{\rm ren}_{\rm{L},k=1} \stackrel{D \gg \tau_k/\Omega}{=} -&\frac{U}{2i\pi D} \frac{\tau^{\rm ren}_{0}}{(T_{\rm K}+i\Omega)/D} \frac{\epsilon_{k=1}}{D} \left[ - 2 i \arctan\left(\frac{T_{\rm K}}{2\Omega}\right) + i\pi  +\ln\left(\frac{T_{\rm K}^2+4\Omega^2}{T_{\rm K}^2}\right)\right],
\label{threnepsonly}
\end{align}
 where $\Gamma_{1d}$ has been replaced by $T_{\rm K}$, incorporating the feedback of $\tau_0$. The marginal renormalization of $\epsilon_{k=1}$ renders it unnecessary to include its feedback. This calculation can be done analogously for the right hopping element $\tau_{{\rm R},k=1}$.

To compute the $\Lambda$ dependent flow of the parameter, we skip the last step in building the effective model and keep the four site model as depicted in the upper right panel of Fig.\,\ref{fig:replicaepsonly}. 
This is necessary to correctly add the auxiliary reservoirs to each of the four sites of the model.
 $\tau^{\Lambda}_{\text{L},k=1}$ is then computed as
\begin{equation}
\tau^{\Lambda}_{\rm{L},k=1} = -\frac{U}{2\pi i} \int d\omega\, G^{<,\Lambda}_{14} (\omega)  ,
\end{equation} 
with $G^{<,\Lambda}_{14}$ set up in the effective model with four sites and the $\omega$ integral is evaluated numerically for each value of $\Lambda$. 

\subsection{Current formulas}
The mean left current in protocol 1 can be described by the following expression for temperatures $T$ smaller than the driving frequency
\begin{align}
J_{{\rm L},k=0}&\stackrel{\phantom{\Omega\gg\epsilon}}{=}\frac{1}{2\pi}\int\limits_{-\infty}^{\infty}dE\;
\left[f(E)-f(E-\Omega)\right]\frac{2}{D}\frac{|\tau_{L,1}|^2 T_K/2}{(E-\epsilon-\Omega)^2+(T_K/2)^2} \nonumber \\
&\phantom{\stackrel{\Omega\gg\epsilon}{=}}+\frac{1}{2\pi}\int\limits_{-\infty}^{\infty}dE\;\left[f(E)-
f(E+\Omega)\right]\frac{2}{D}\frac{|\tau_{L,1}|^2 T_K/2}{(E-\epsilon+\Omega)^2+(T_K/2)^2} \hspace{3cm}
\label{eq:curfinT}
\end{align} 
for $\epsilon\ll\Omega\ll D$ and where $T_K= 4\tau_0^2/D$. 
For $T=0$ this simplifies to
\begin{align}
J_{{\rm L},k=0}&\stackrel{\phantom{\Omega\gg\epsilon}}{=}\frac{1}{2\pi}\int\limits_{-\infty}^{\infty}dE\;
\left[\Theta(-E)-\Theta(-E+\Omega)\right]\frac{2}{D}\frac{|\tau_{L,1}|^2 T_K/2}{(E-\epsilon-\Omega)^2+(T_K/2)^2} \nonumber \\
&\phantom{\stackrel{\Omega\gg\epsilon}{=}}+\frac{1}{2\pi}\int\limits_{-\infty}^{\infty}dE\;\left[\Theta(-E)-
\Theta(-E-\Omega)\right]\frac{2}{D}\frac{|\tau_{L,1}|^2 T_K/2}{(E-\epsilon+\Omega)^2+(T_K/2)^2}\nonumber \\
&\stackrel{\phantom{\Omega\gg\epsilon}}{=}\frac{1}{2\pi}\left(\frac{|\tau_{L,1}|}{\tau_{0}}\right)^2 T_K\left[\arctan\left(\frac{2\epsilon}{T_K}\right)+\frac12\arctan\left(\frac{2\Omega-2\epsilon}{T_K}\right)-\frac12\arctan\left(\frac{2\Omega+2\epsilon}{T_K}\right)\right]\nonumber \\
&\stackrel{\Omega\gg\epsilon}{=}\frac{1}{2\pi}\left(\frac{|\tau_{L,1}|}{\tau_{0}}\right)^2 T_K\arctan\left(\frac{2\epsilon}{T_K}\right).
\label{eq:cur}
\end{align} 
For more details on the derivation see the Supplementary Material of Ref.\onlinecite{Eissing16}.

\subsection{Master equation in Floquet Space}
We consider the noninteracting model, which simplifies to
\begin{equation}
H = \epsilon d^{\dagger} d + \sum_{\alpha, q_{\alpha}}  \left[ \epsilon_{q_{\alpha}} a^{\dagger}_{q_{\alpha}} a_{q_{\alpha}} + \sqrt{\frac{T_{\rm K}}{4\pi \nu}} \left( d a^{\dagger}_{q_{\alpha}} +d^{\dagger} a_{q_{\alpha}}  \right) \right]
\end{equation}
with $\nu = \sum_{q_{\alpha}} \delta(\omega - \epsilon_{q_{\alpha}})$ and the non-interacting $T_{\rm K}$ and employ the wideband limit.\newline
The kinetic equation as well as the current formula in the Liouville space are
\begin{align}
i \dot{ \boldsymbol{\rho}} (t) &=\int_{t_0}^t dt' \mathbf{L} (t,t') \boldsymbol{\rho} (t'), \\
\left< I_\gamma\right> (t) &= -i {\rm Tr} \left(\int_{t_0}^t \boldsymbol{\Sigma}_\gamma (t,t') \boldsymbol{\rho} (t')\right),
\end{align}
with the Liouvillian $\mathbf{L}(t,t') = \mathbf{L}_S (t,t') + \boldsymbol{\Sigma} (t,t')$, the reduced density matrix of $\boldsymbol{\rho} (t')$ and the system Liouvillian $\mathbf{L}_S$, where all reservoir degrees of freedom have been traced out following the standard approach.\cite{Schoeller2009}
Here we employ the approximation of separating time scales and set $\boldsymbol{\rho} (t') \rightarrow \boldsymbol{\rho} (t)$. Since we are interested in the long time behavior of a time periodic system, we can rewrite the selfenergy or the current kernel as
\begin{equation}
 \boldsymbol{\Sigma}(t,t') = \sum_n e^{-in\Omega t}  \boldsymbol{\Sigma}_n (t-t') ,
\end{equation}
which allows to rewrite the two equations to
\begin{align}
i \dot{ \boldsymbol{\rho}} (t) &=\sum_k e^{-ik\Omega t}  \boldsymbol{\Sigma}^k (i0^+) \boldsymbol{\rho} (t) , \\
\left< I_\gamma\right> (t) &= -i {\rm Tr} \left(\sum_k e^{-ik\Omega t}  \boldsymbol{\Sigma}_\gamma^k (i0^+) \boldsymbol{\rho} (t)\right) .
\end{align}

\phantom{x}
\noindent Due to the charge conservation in this simple quantum dot model, the off diagonal blocks vanish and it is sufficient to calculate the probabilities of the dot being empty or full ($p_0,p_1$). The equation for the probabilities reduces to a quantum master equation in Floquet space
\begin{align}
 \dot{p}_s (t) &=\sum_{ks'} e^{-ik\Omega t} W^k_{ss'} p_{s'} (t), \nonumber \\
&=\sum_{k n s'\neq s} e^{-i(k+n)\Omega t} ( W^k_{ss'} p^n_{s'} - W^k_{s's} p^n_s) \label{eq:kin}\\
\left< I_\gamma\right> (t) &=  \sum_{knss'} e^{-i(k+n)\Omega t} W^{k,\gamma}_{ss'} p^n_{s'} ,
\end{align}
where we have defined $-i\Sigma^{k,(\gamma)}_{ss,s's'} (i0^+) = W^{k,(\gamma)}_{ss'}$ and $s,s' \in \lbrace0,1\rbrace$.
The entries of the (current) kernel are
\begin{align}
W^k_{00} &= \phantom{-} \Gamma^k_0  &&\hspace{1cm} W^{k,\alpha}_{00} = - \Gamma^{k,\alpha}_0 /2 &&\nonumber \\
W^k_{01} &= - \Gamma^k_1 &&\hspace{1cm}  W^{k,\alpha}_{01} = \phantom{-} \Gamma^{k,\alpha}_1 /2 &&\nonumber \\
W^k_{10} &= - \Gamma^k_0 &&\hspace{1cm}  W^{k,\alpha}_{10} = - \Gamma^{k,\alpha}_0 /2 &&\nonumber \\
W^k_{11} &= \phantom{-}\Gamma^k_1  &&\hspace{1cm}  W^{k,\alpha}_{11} = \phantom{-} \Gamma^{k,\alpha}_1/2 &&\nonumber 
\end{align}
with $ \Gamma^k_{0/1}  =  \Gamma^{k,{\rm L}}_{0/1}+  \Gamma^{k,{\rm R}}_{0/1}$.

The kernel $W^k$ is computed to the first order in the tunneling rate in Floquet space.
We only consider setups in which the hoppings are time periodic. In addition to the known diagrammatical rules,\cite{Schoeller2009, Kashuba2013} each time dependent hopping vertex aquires a Floquet index and the energy argument of the propagator is shifted by $(k-\sum_j j)\Omega$ where $j$ runs over all Floquet indices of hopping vertices to the left.
Furthermore, a $\delta_{k,\sum_j}$ is required. The rates then compute to
\begin{align}
\Gamma_0^k
&= \sum_{\alpha, k_2} t^{\alpha}_{k-k_2} t^{\alpha}_{k_2} \left[  \sum_{n \geq 0} \frac{T_\alpha\pi}{k_2\Omega+ \epsilon+i\omega_{n}^{\alpha}}-  \sum_{n \geq 0} \frac{T_\alpha\pi}{k_2\Omega- \epsilon+i\omega_{n}^{\alpha}} -\pi\right], \nonumber \\
\Gamma_1^k &= \sum_{\alpha, k_2}  t^{\alpha}_{k-k_2} t^{\alpha}_{k_2} \left[ \sum_{n \geq 0} \frac{T_\alpha\pi}{k_2\Omega- \epsilon+i\omega_{n}^{\alpha}}- \sum_{n \geq 0} \frac{T_\alpha\pi}{k_2\Omega+ \epsilon+i\omega_{n}^{\alpha}}-\pi\right], \nonumber \\ 
\end{align}
with temperature $T_{\alpha}$ of reservoir $\alpha \in (\rm L,R)$ and $\omega_{n}^{\alpha} = (2n+1)\pi T_{\alpha}$ are the respective fermionic Matsubara frequencies. 

In order to solve Eq.(\ref{eq:kin}) for the time periodic steady state, we can diagonalize the kernel $W^k$ and use the spectral decomposition. Only the eigenvector to the eigenvalue $\lambda = 0$  is necessary to calculate the steady state of $\rho(t)$ in the long time limit. Equivalently, Eq.(\ref{eq:kin}) can be rewritten under the assumption of a time periodic form of the probability as well as by employing the symmetries $ p_0^0 = 1-p_1^0 ; p_0^n = -p_1^n $, which leads to coupled expressions for the higher harmonics
\begin{align}
p^0_0 &= \frac{1}{\Gamma^0_1 +\Gamma^0_0} (\Gamma_1^0- \sum_{k \neq 0} (\Gamma_1^k + \Gamma_0^k) p_0^{-k}), \\
p^0_1 &= \frac{1}{\Gamma^0_1 +\Gamma^0_0} (\Gamma_0^0- \sum_{k \neq 0} (\Gamma_1^k + \Gamma_0^k) p_1^{-k}), \\
p^m_0 &= - \frac{1}{-im\Omega + \Gamma^0_1 +\Gamma^0_0}\left(\sum_{k \neq 0} (\Gamma_1^k + \Gamma_0^k) p_0^{m-k} + \Gamma_0^m\right)\hspace{1cm} m \neq 0, \\
p^m_1 &= - \frac{1}{-im\Omega + \Gamma^0_1 +\Gamma^0_0}\left(\sum_{k \neq 0} (\Gamma_1^k + \Gamma_0^k) p_1^{m-k} + \Gamma_1^m\right) \hspace{1cm} m \neq 0 .
\end{align}

\phantom{x}
\noindent The left mean current can be rewritten as
\begin{align}
\left< I_L\right> (t) &= \sum_{nss'}  W^{-n,\gamma}_{ss'} p^n_{s'} \nonumber \\
&= \sum_n \Gamma^{-n,{\rm L}}_{1} p_1^n - \Gamma^{-n,{\rm L}}_{0} p_0^n .
\end{align}

\twocolumngrid

\bibliographystyle{apsrev4-1}
\bibliography{LongFloquetFRG}
\end{document}